\documentclass[11pt]{article}
\usepackage{caption}

\usepackage{cite}

\usepackage{subcaption}
\usepackage[utf8]{inputenc}

\usepackage{amssymb}
\usepackage{amsfonts}
\usepackage{amsmath}
\usepackage{graphicx}
\usepackage{indentfirst}
\usepackage{dsfont}

\usepackage{mathtools}

\usepackage{enumerate}
\usepackage{url}
\usepackage{lmodern}
\usepackage{newpxtext}

\usepackage{amsthm}

\usepackage{tikz}
\definecolor{darkblue}{rgb}{0.15,0.35,0.55}
\definecolor{reddish}{rgb}{.8, 0.2, 0.2}

\long\def\ca#1\cb{} 

\newcommand{\becs}{\begin{cases}}
\newcommand{\bem}{\begin{matrix}}

\newcommand{\bra}[1]{\langle#1|}

\newcommand{\bsk}{\bigskip }

\newcommand{\dya}[1]{|#1\rangle\langle#1|}
\newcommand{\dyad}[2]{|#1\rangle\langle#2|}

\newcommand{\encs}{\end{cases}}
\newcommand{\enm}{\end{matrix}}

\newcommand{\inp}[1]{\langle#1|#1\rangle }
\newcommand{\inpV}[2]{\langle#1,#2\rangle }
\newcommand{\inpd}[2]{\langle#1|#2\rangle }
\newcommand{\ket}[1]{|#1\rangle }

\newcommand{\mat}[1]{\left(\begin{matrix}#1\end{matrix}\right)}

\newcommand{\ot}{\otimes }

\newcommand{\Tr}{{\rm Tr}}

\newcommand{\BC}{{\mathcal B}}
\newcommand{\CC}{{\mathcal C}}
\newcommand{\DC}{{\mathcal D}}
\newcommand{\EC}{{\mathcal E}}
\newcommand{\FC}{{\mathcal F}}

\newcommand{\HC}{{\mathcal H}}
\newcommand{\IC}{{\mathcal I}}
\newcommand{\JC}{{\mathcal J}}

\newcommand{\LC}{{\mathcal L}}
\newcommand{\MC}{{\mathcal M}}
\newcommand{\NC}{{\mathcal N}}

\newcommand{\QC}{{\mathcal Q}}

\newcommand{\SC}{{\mathcal S}}
\newcommand{\TC}{{\mathcal T}}

\newcommand{\XC}{{\mathcal X}}
\newcommand{\YC}{{\mathcal Y}}

\newcommand{\rB}{\textbf{r}}

\newcommand{\vB}{\textbf{v}}
\newcommand{\wB}{\textbf{w}}

\newcommand{\Rbb}{\mathbb{R}}

\newcommand{\al}{\alpha }

\newcommand{\gm}{\gamma }
\newcommand{\Gm}{\Gamma }
\newcommand{\dl}{\delta }
\newcommand{\Dl}{\Delta }
\newcommand{\ep}{\epsilon}

\newcommand{\lm}{\lambda }
\newcommand{\Lm}{\Lambda }

\newcommand{\sg}{\sigma }

\usepackage[symbol]{footmisc}

\renewcommand{\thefootnote}{\fnsymbol{footnote}}

\usepackage[pdfpagelabels, linktocpage=true]{hyperref}
\hypersetup{
colorlinks=true,
citecolor=darkblue,
linkcolor=reddish,
urlcolor=darkblue,
pdfauthor={},
pdftitle={},
pdfsubject={}
}

\begin{document}

 \begin{center}
{\bf Five Starter Pieces: Quantum Information Science via Semi-definite Programs}\bsk\\
     \normalsize Vikesh Siddhu~$^1$~\footnote{\href{mailto:vsiddhu@protonmail.com}{\texttt{vsiddhu@protonmail.com}}}
     and Sridhar Tayur~$^2$~\footnote{\href{mailto:stayur@cmu.edu}{\texttt{stayur@cmu.edu}}}\\
    \textit{$^1$ JILA, University of Colorado/NIST, 440 UCB, Boulder, CO 80309, USA}\\
    \textit{$^2$ Quantum Technologies Group, Tepper School of Business, Carnegie Mellon University, Pittsburgh PA 15213, USA}\\
Date: October 18, 2022
\end{center}

\renewcommand*{\thefootnote}{\arabic{footnote}}
\setcounter{footnote}{0}

\begin{abstract}
    As the title indicates, this chapter presents a brief, self-contained
    introduction to five fundamental problems in Quantum Information Science
    (QIS) that are especially well-suited to be formulated as Semi-definite
    Programs (SDP). We have in mind two audiences. 
    The primary audience comprises of Operations Research (and Computer
    Science) graduate students who have familiarity with SDPs, but have found
    it daunting to become even minimally conversant with pre-requisites of QIS.
    The second audience consists of Physicists (and Electrical Engineers)
    already knowledgeable with modeling of QIS via SDP but interested in
    computational tools that are applicable more generally. 
    For both audiences, we strive for rapid access to the unfamiliar material.
    For the first, we provide just enough required background material (from
    Quantum Mechanics, treated via matrices, and mapping them in Dirac
    notation) and simultaneously for the second audience we recreate,
    computationally in Jupyter notebooks, known closed-form solutions.
    We hope you will enjoy this introduction and gain understanding of the
    marvelous connection between SDP and QIS by self-study, or as a short
    seminar course.  Ultimately, we hope this disciplinary outreach will fuel
    advances in QIS through their fruitful study via SDPs.
\end{abstract}


\tableofcontents

\newpage

\section{Introduction}

Thomas Sprat, in 1667, as historian at the Royal Society of London, noted a
connection between being an outsider to a trade and inventiveness:
\begin{quote}
    A glance from an angle might well reveal a new aspect of nature. 
\end{quote}
We would like to create such a trading zone through this chapter - and
invite Operations Researchers and Computer Scientists - to foster innovative
contributions to Quantum Information Science~(QIS). 

QIS spans a variety of sub-fields including quantum computing and quantum
communication~\cite{NielsenChuang11}. Quantum computing offers a novel way to
perform calculations which could be faster than regular~(classical) computing
for several important problem classes, such as prime
factorization~\cite{Shor97}.  This novelty and speed comes from utilizing
properties of quantum mechanics such as superposition and
entanglement~\cite{HorodeckiHorodeckiEA09a}.  
Quantum communication can not only carry a new type of information using qubits
("quantum bit"), it can also be used to communicate regular~(classical)
information (bit) with greater privacy~\cite{BennettBrassard84,BennettShor98}.
Furthermore, quantum communication can be non-additive: two quantum
communication devices can send more together than each device used
separately~\cite{SmithYard08}. 
These potential advantages of using quantum states for information processing
are often abated by noise. This noise affects quantum states and their
properties important for computation and communication. 
To fully understand and leverage quantum technologies computation and
communication, it is important to study basic properties of quantum states such
as entanglement and information theoretic properties such as capacity to carry
information~\cite{BennettShor04}.
In such studies semi-definite programs~(SDPs) play a useful role. Since SDPs
are a well-known tool in Operations Research~(OR) and Computer Science~(CS),
they offer researchers in OR and CS communities a natural way to interact with
QIS.

Semi-definite programs~(for a brief introduction,
see~\cite{VandenbergheBoyd96, OvertonWolkowicz97, Todd01}) are an extension of
linear programs~(LPs) obtained by replacing element wise non-negative vector
variables in LPs with positive semi-definite matrices. This replacement results
in a non-linear, but convex optimization problem, which is much more general
than an LP. However, this general SDP carries with it a variety of nice
properties of LPs which make it possible to efficiently solve SDPs, both in
theory and practice. For instance, most of the theory of duality directly
extends from LPs to SDPs~\cite{BellmanFan63}. The simplex method for
LPs~\cite{DantzigThapa97, Schrijver98} can, \textcolor{black}{in principle}, be extended to SDPs~\cite{Pataki96}.
For SDP constraints, one can construct cutting planes in polynomial
time~\cite{BoydElGhaouiEA94} and thus use a polynomial-time ellipsoid
method~\cite{NemirovskyYudin77, Shor77} to numerically solve an SDP. However,
in practice it is often faster to use interior point methods~\cite{Alizadeh92,
NesterovNemirovskii94, KamathKarmarkar14}~(such as those extending Karmarkar's
interior point method~\cite{Karmarkar84} for LPs) to efficiently solve SDPs.

The ability to efficiently solve SDPs is not their only draw. These
optimization programs naturally appear in a variety of fields including control
theory~\cite{BoydElGhaouiEA94}, graph theory~\cite{GoemansWilliamson95,
KargerMotwaniEA98}, combinatorial optimization~\cite{Alizadeh95}, and algebraic
geometry~\cite{HauensteinLiddellEA21}. SDPs in these and other engineering
fields usually have real positive semi-definite matrix variables.  {\em
Complex} positive semi-definite matrices naturally appear, and play an
important role in quantum mechanics, a linear theory in the physical sciences
which successfully describes the physical world. It is no surprise that a
variety of fundamental and applied problems in quantum theory can be re-written
as SDPs~\cite{Watrous18, Wang18,KhatriWilde20}. 
Such re-writing has been fueled by the growth of quantum computing and information
science, which study and hope to practically perform information processing using
physical objects accurately described by quantum theory. 
Quantum information science offers an exciting and potentially fertile area
where SDPs and other optimization techniques can continue to play an important
role. Standard exposition of quantum mechanics often involves new notation,
unitary dynamics, and other historical aspects of quantum theory.  This
route to learning quantum mechanics has its advantages, but it can create a
barrier to entry for optimization experts working outside the area of quantum
information science~(QIS). The key motivation for this work is to lower this
barrier and expose a broader audience to the recent SDP work in QIS.

In what follows, in Secs.~\ref{SQMBare} and~\ref{SQIBare} we provide a bare
bones introduction to quantum mechanics and information theory with running
examples, most using $2 \times 2$ matrices. In this introduction we not only
cover basic concepts likes quantum states, measurements, Born's rule~(see
Sec.~\ref{BornRule}), entanglement, entropy, and quantum channels, but also
take the opportunity to introduce {\em Dirac notation}, which is standard
across quantum theory and QIS. Next we present five problems in QIS: quantum
state discrimination~(in Sec.~\ref{P1StDis}), state fidelity~(in
Sec.~\ref{P2StFid}), channel discrimination~(in Sec.~\ref{P3ChDis}),
entanglement and separability~(in Sec.~\ref{P4EntSep}), and channel
capacity~(in Sec.~\ref{P5Cap}).  These problems, presented in order of
increasing level of mathematical sophistication, by no means comprise a complete
list of problems in QIS where SDPs and other optimization techniques are of
value. However, they offer a strong stepping stone to continue future
exploration of this type.
For each problem, we provide a motivation, a crisp mathematical statement, an
SDP formulation, certain special cases~(sometimes with algebraic SDP
solutions), numerical examples with working Python notebooks, and avenues for
future exploration.

\section{Quantum Mechanics: The Bare Minimum}
\label{SQMBare}

\subsection{Quantum states and Dirac notation}
\label{SStatesDirac}

Quantum mechanics accurately predicts measurable properties of microscopic
physical objects. While these physical objects live in the real world, they are
described in quantum mechanics using complex numbers. The simplest complex
number, $i$, is just the square root of minus one. In general, any complex
number can be written as $z=x + iy$, where $x,y$ are real numbers; that is, $x,y
\in \mathbb{R}$. We say $z \in \mathbb{C}$, the space of complex numbers. The
complex conjugate of $z$ is $x -iy$ and is represented by $z^*$. Using $z$ and
$z^*$ one constructs the norm, $|z| = \sqrt{z^*z} = \sqrt{x^2 + y^2}$, of a
complex number.

Of main interest in quantum mechanics are tuples of complex numbers.  A length
$d$ tuple of this type is just a column vector $\vB$ in $d$-dimensional complex
space $\mathbb{C}^d$. The inner product of a column vector $\vB$ with $\wB$,
$\vB^* \cdot \wB$, resembles the ordinary dot product $\vB \cdot \wB$ of real vectors, except
the entries of the first column vector $\vB$ are complex conjugated. A
collection of $d$ column vectors, $\{\vB_i\}$, where each vector has unit
norm~($\vB_i^*\vB_i=1$) and any distinct pair of vectors are orthogonal, that is,
the inner product, $\vB_i^*\vB_j = 0$, for $i \neq j$, is called an {\em
orthonormal basis} of $\mathbb{C}^d$. Using this orthonormal basis, any vector
in $\mathbb{C}^d$ can be written as a linear combination, $\sum_i c_i \vB_i$,
where $c_i \in \mathbb{C}$.

In quantum physics literature, the space $\mathbb{C}^d$, its column vectors
$\vB$, and the inner product of a column $\vB$ with another column vector $\wB$
are denoted by $\HC$, $\ket{v}$, and $\inpd{v}{w}$, respectively. This notation
is called {\em Dirac} notation, where $\inpd{v}{w}$ is called a {\em braket},
its first half, $\bra{v}$, is called a {\em bra} and the second half, $\ket{w}$
is called a {\em ket}. The ket $\ket{v}$ is represented by a column vector.
Taking the transpose of this column vector and then complex conjugating each
entry results in a row vector. This row vector represents the bra $\bra{v}$. The
multiplication of a ket $\ket{v}$ by a scalar $c \in \mathbb{C}$ is denoted as
$c \ket{v}$.
\textcolor{black}{Any ket $\ket{v}$ that has unit inner product with itself,
$\inpd{v}{v} = 1$, is called a {\em pure state}}.

The inner product $\inpd{v}{w}$ is a complex number obtained by multiplying the
row vector $\bra{v}$ with the column vector $\ket{w}$. By interchanging the
order of multiplication, we multiply a $d$-dimensional column vector $\ket{w}$
with a $d$-dimensional row vector $\bra{v}$ to obtain the outer-product,
$\dyad{w}{v}$, which is a $d \times d$ square matrix with complex entries. This
square matrix represents a linear operator. We denote the set of linear
operators on $\HC$ by $\LC(\HC)$.  The action of the operator $\dyad{w}{v}$ on $\ket{u} \in
\HC$ is given by 
\begin{equation}
    (\dyad{w}{v} )\ket{u} = 
    \ket{w} (\inpd{v}{u}) = (\inpd{v}{u}) \ket{w}.
    \label{dyadAction}
\end{equation}
The equality on the right technically defines the action of the operator on the
left. However, the middle term, obtained by removal of a parenthesis and
replacement of two vertical bars $||$ between $v$ and $u$ with one bar $|$ is
an example of slickness embedded in Dirac notation. This slickness explains the
action of operators without doing a matrix calculation. For instance, the
result in~\eqref{dyadAction} is essentially a matrix calculation where the $(d
\times d)$ matrix for $\dyad{w}{v}$ is multiplied with a $d$-dimensional column
vector $\ket{u}$, to obtain the outcome, $(\inpd{v}{u}) \ket{w}$.

Unlike $\dyad{w}{v}$, not all linear operators are dyads, linear operators
can be written as sums of dyads and represented by matrices. 
The transpose of a matrix $M$ is denoted by
$M^{T}$, and the adjoint $M^{\dag}$ is obtained by complex conjugating each
entry of $M^T$~(see footnote~\footnote{\textcolor{black}{We have used notation which
is common in physics where $*$ and $\dag$ denote complex conjugate and adjoint
operations, respectively. In mathematics and optimization, it is common to use
$\bar{z}$ for complex conjugate of $z$ and $*$ for adjoint operation.  There is
yet another combination, $\bar{z}$ for complex conjugate of $z$ but $\dag$ for
adjoint operations, which can be seen in some physics, computer science and
optimization literature.}} for commment on the notation). 
If $N$ is another linear operator, then $(NM)^{\dag} = M^{\dag}N^{\dag}$.
For square matrices, those with equal numbers of rows and columns, we
denote the matrix determinant by $\det(M)$. Square matrices with non-zero
determinants can be inverted, and the matrix inverse $M^{-1}$ satisfies
$MM^{-1} = M^{-1}M = I$, where $I$ is the identity matrix. For any square
matrix $M$, there is a set of non-zero vectors $\{\ket{a_i}\}$
such that $M$ satisfies 
\begin{equation}
    M \ket{a_i} = \al_i \ket{a_i}.
    \label{eigen}
\end{equation}
Here complex numbers $\al_i$ are called eigenvalues and $\ket{a_i}$ are called
eigenvectors.  A square matrix $M$ which commutes with its adjoint, $MM^{\dag}
= M^{\dag}M$, is called a normal matrix. Any normal matrix $M$ can be
diagonalized using an orthonormal basis,
\begin{equation}
    M = \sum_i \al_i \dya{m_i},
    \label{eigDec}
\end{equation}
where the basis vector $\ket{m_i}$ is an eigenvector of $M$ and has the
complex eigenvalue $\al_i$. Two special types of normal matrices are of
particular interest in quantum mechanics. One is a unitary matrix, usually
denoted by $U$, which satisfies $UU^{\dag} = U^{\dag}U = I$; another is a
Hermitian matrix, $O$, which satisfies $O = O^{\dag}$.
Before proceeding forward, we illustrate the use of the adjoint operation
$\dag$ in Dirac notation.  Suppose $\ket{\psi}$ is any ket, then
$\ket{\psi}^{\dag} = \bra{\psi}$ and $\bra{\psi}^{\dag} = \ket{\psi}$. If
$\ket{\phi}$ is another ket, then $(\dyad{\psi}{\phi})^{\dag} =
\dyad{\phi}{\psi}$. For complex numbers $c_0$ and $c_1$, the adjoint of the
linear combination $(c_0 \ket{\psi} + c_1 \ket{\phi})^{\dag} = c_0^* \bra{\psi}
+ c_1^* \bra{\phi}$. If $N$ is a linear operator, then $\bra{i}N^{\dag}\ket{j}
= (\bra{j}N \ket{i})^*$.

The simplest non-trivial space $\HC$ has dimension two~($d=2$); that is,
any $\ket{\psi} \in \HC$ can be written as a linear combination of two
orthonormal vectors. It is customary to introduce a {\em standard}~(or {\em
computational}) basis $\{ \ket{0}, \ket{1}\}$ for $\HC$ with $d = 2$. Here
$\inp{0} = \inp{1}= 1$ and $\inpd{0}{1} = 0$. It is common
to represent the computational basis as column vectors
\begin{equation}
    \ket{0} := \begin{pmatrix} 1 \\ 0 \end{pmatrix}, \quad \text{and} \quad
    \ket{1} := \begin{pmatrix} 0 \\ 1 \end{pmatrix}.
    \label{computeBasis}
\end{equation}
As mentioned earlier, the inner product $\inpd{0}{1}$ can \textcolor{black}{be} obtained by
multiplying each row of $\ket{1}$ with the complex conjugate of each row of
$\ket{0}$.
In general, the state of a two-dimensional quantum system, called a {\em qubit},
is given by $\ket{\psi} = c_0 \ket{0} + c_1 \ket{1}$ where $|c_0|^2 + |c_1|^2 =
1$; $\ket{\psi}$ can be written as a column vector
\begin{equation}
    \ket{\psi} := \begin{pmatrix} c_0 \\ c_1 \end{pmatrix}.
    \label{qbitCol}
\end{equation}
The notation $\ket{0}$ and $\ket{1}$ is intended to draw an analogy with
classical bits. Just like the distinguishable states $0$ and $1$ of a bit, the
quantum states $\ket{0}$ and $\ket{1}$ represent perfectly distinguishable
states of a qubit. Like any classical analogy for a quantum system, this
analogy between bits and qubits has its limitations.  For instance, the linear
combination $c_0 \ket{0} + c_1 \ket{1}$, where $|c_0|^2 + |c_1|^2 = 1, |c_0|
>0,$ and $|c_1|>0$, is a perfectly well-defined quantum state; however, there is
no analogous state of a classical bit.
There are particular linear combinations of the standard basis elements that
are of special interest.  One such linear combination is 
\begin{equation}
    \ket{+} := \frac{1}{\sqrt{2}}(\ket{0} + \ket{1}) 
    = \frac{1}{\sqrt{2}} \begin{pmatrix} 1 \\ 1 \end{pmatrix}, 
    \quad \text{and} \quad
    \ket{-} := \frac{1}{\sqrt{2}}(\ket{0} - \ket{1}) 
    = \frac{1}{\sqrt{2}} \begin{pmatrix} 1 \\ -1 \end{pmatrix}.
    \label{hdmBasis}
\end{equation}
Notice, $\inp{+} = \inp{-}= 1$ and $\inpd{+}{-} = 0$; as a result $\{ \ket{+},
\ket{-}\}$, forms a basis of $\HC$. This basis is sometimes called the {\em
Hadamard basis} because it can be obtained from the computational basis,
$\ket{+} = H \ket{0}$ and \textcolor{black}{$\ket{-} = H \ket{1}$}, using the Hadamard
operator 
\begin{equation}
    H = \frac{1}{\sqrt{2}}
    \begin{pmatrix}
        1 & 1 \\
        1 & -1
    \end{pmatrix}.
    \label{hMatrix}
\end{equation}
Notice the Hadamard matrix $H$ is unitary~($HH^{\dag} = H^{\dag}H = I$) and
Hermitian~(i.e., $H = H^{\dag}$); as a consequence $H^2 = I$.

Let $\{\ket{j}\}$ be the computational basis of a $d$-dimensional space $\HC$.
In Dirac notation, the identity operator $I$ on $\HC$ can be written as
\begin{equation}
    I = \sum_{j=0}^{d - 1} \dya{j}.
    \label{idOp}
\end{equation}
The expression above is often called the {\em completeness relation}. This
relation can be useful. For instance, suppose $\ket{\psi}$ is any ket in
$\HC$, represented by some column vector. One may use the completeness
relation as follows:
\begin{equation}
    \ket{\psi} = I \ket{\psi} 
    = \sum_j (\dya{j}) \ket{\psi} 
    = \sum_j \inpd{j}{\psi} \ket{j},
\end{equation}
to find that $\inpd{j}{\psi}$ is simply the $j^{\text{th}}$ entry~($j$ starts
from zero) of the column vector representing $\ket{\psi}$ in the standard
basis. For instance, the column vector $\ket{\psi}$ in~\eqref{qbitCol} has $c_0
= \inpd{0}{\psi}$ and $c_1 = \inpd{1}{\psi}$.
The completeness relation can also be used to find entries of a linear operator
$N$ on $\HC$,
\begin{equation}
    N = I N I = \big( \sum_i \dya{i} \big) N \big( \sum_j \dya{j} \big) 
    = \sum_{i,j} \bra{i} N \ket{j} \dyad{i}{j}
\end{equation}
to find that the $i^{\text{th}}$ row and $j^{\text{th}}$ column of $N$ is simply
$\bra{i} N \ket{j}$. If $d$ is two, then the matrix for $N$ in the standard
basis is simply
\begin{equation}
    N = \begin{pmatrix}
        \bra{0} N \ket{0} &  \bra{0} N \ket{1} \\ 
        \bra{1} N \ket{0} &  \bra{1} N \ket{1} 
    \end{pmatrix}.
\end{equation}
In general, the trace of an operator $N$, $\Tr(N)$, is simply $\sum_{j} \bra{j}
N \ket{j}$.  Uses of the completeness relation, similar to those above, can show
that $\Tr(\dyad{\psi}{\phi}) = \inpd{\phi}{\psi}$ and $\Tr(A \dyad{\psi}{\phi})
= \bra{\phi} A \ket{\psi}$.

So far we have focused on a single quantum system. Suppose there are two
systems $a$ and $b$ in quantum states $\ket{v}_a \in \HC_a$ and $\ket{w}_b \in
\HC_b$, respectively. Then state of the joint $ab$ system is written as
$\ket{v}_a \ot \ket{w}_b$, a tensor product, $\ot$, of $\ket{v}_a$ and
$\ket{w}_b$. This joint state belongs to a space $\HC_{ab} = \HC_a \ot \HC_b$
obtained by taking the tensor product of $\HC_a$ with $\HC_{\textcolor{black}{b}}$.
Suppose $\ket{v}_a$ and $\ket{w}_b$ are represented by column vectors of
dimension $d_a$ and $d_b$, respectively; then $\ket{v}_a \ot \ket{w}_b$ is
given by a column vector of dimension $d_a d_b$ formed by taking the {\em
Kronecker product} of each column vector.  For instance, let $d_a = d_b = 2$,
\begin{equation}
    \ket{v}_a = \begin{pmatrix} c_0 \\ c_1 \end{pmatrix},\quad
    \ket{w}_b = \begin{pmatrix} d_0 \\ d_1 \end{pmatrix}, 
        \label{tenProd1}
\end{equation}
where $|c_0|^2 + |c_1|\textcolor{black}{^2} = |d_0|^2 + |d_1|^2 = 1$, then
\begin{equation}
    \ket{v}_a \ot \ket{w}_b = 
    \begin{pmatrix} 
        c_0 \begin{pmatrix} d_0 \\ d_1 \end{pmatrix} \\
        c_1 \begin{pmatrix} d_0 \\ d_1 \end{pmatrix} 
    \end{pmatrix}
    \quad = 
    \begin{pmatrix} 
        c_0 d_0 \\ c_0 d_1 \\ c_1 d_0 \\c_1 d_1
    \end{pmatrix}.
        \label{tenProd2}
\end{equation}
In general, the state $\ket{\psi} \in \HC_{ab}$ can be written as a linear
combination of an orthonormal basis of $\HC_{ab}$. One simple orthonormal basis
of this type can be constructed by taking tensor products of the computational
basis elements of $\HC_a$ and $\HC_b$, respectively. For instance, if $d_a =
2$, $d_b = 2$, $\{ \ket{0}_a, \ket{1}_a\}$ and $\{\ket{0}_b, \ket{1}_b \}$ are
computational basis of $\HC_a$ and $\HC_b$, respectively, then $\{\ket{0}_a \ot
\ket{0}_b, \ket{0}_a \ot \ket{1}_b, \ket{1}_a \ot \ket{0}_b, \ket{1}_a \ot
\ket{1}_b \}$ is an orthonormal basis for $\HC_{ab}$.  This basis can be
represented as follows
\begin{equation}
    \ket{0}_a \ot \ket{0}_b = 
    \begin{pmatrix} 
        1 \\ 0 \\ 0 \\ 0
    \end{pmatrix},
    \quad
    \ket{0}_a \ot \ket{1}_b = 
    \begin{pmatrix} 
        0 \\ 1 \\ 0 \\ 0
    \end{pmatrix},
    \quad
    \ket{1}_a \ot \ket{0}_b = 
    \begin{pmatrix} 
        0 \\ 0 \\ 1 \\ 0
    \end{pmatrix},
    \quad \text{and} \quad
    \ket{1}_a \ot \ket{1}_b = 
    \begin{pmatrix} 
        0 \\ 0 \\ 0 \\ 1
    \end{pmatrix}.
\end{equation}
States of the two qubit system $\HC_{ab}$ can be written as a linear
combination of the basis above.  A simple linear combination of the basis
elements above is
\begin{equation}
    \ket{\phi} = \frac{1}{2} (\sum_{i,j} \ket{i}_a \ot \ket{j}_b).
\end{equation}
It turns out that $\ket{\phi}$ can be written as $\ket{+}_a \ot \ket{+}_b$ and
represents the state of two qubits, each in the state $\ket{+}$. Another
simple linear combination is
\begin{equation}
    \ket{\chi} = \frac{1}{\sqrt{2}}  (\ket{0}_a \ot \ket{0}_b + \ket{1}_a \ot \ket{1}_b ).
    \label{maxEnt}
\end{equation}
Unlike $\ket{\phi}$, the linear combination above cannot be written as
$\ket{v}_a \ot \ket{w}_b$ for any $\ket{v}_a$ and $\ket{w}_b$. Thus, the joint
system $ab$ is in a state that cannot be adequately described by specifying the
state of each individual system $a$ and $b$. Such joint states are called {\em
entangled}. Entanglement is a key aspect of quantum theory. In general, a state
$\ket{\psi}_{ab}$, given by a linear combination $M_{ij} \ket{i}_a \ot
\ket{j}_b$, is entangled if the matrix $M$, with entries $M_{ij}$, has rank
greater than one.

Given two matrices, $N$, mapping $\HC_a$ to itself, and $M$, mapping $\HC_b$ to
itself, one can define their tensor product, $N \ot M$, a matrix from
$\HC_{a} \ot \HC_b$ to itself which acts as follows: 
\begin{equation}
    N \ot M (\ket{v}_a \ot \ket{w}_b) = N \ket{v}_a \ot M \ket{w}_b.
    \label{tenProd3}
\end{equation}
If square matrices $N$ and $M$ have dimensions $d_a$ and $d_b$, respectively,
then the square matrix $N \ot M$ has dimension $d_a \times d_b$. This larger
square matrix is obtained by taking a Kronecker product of $N$ and $M$. For
instance, let $d_a = d_b = 2$, 
\begin{equation}
    N = 
    \begin{pmatrix} 
        N_{00} & N_{01} \\ 
        N_{10} & N_{11}
    \end{pmatrix}, \quad \text{and} \quad
    M = 
    \begin{pmatrix} 
        M_{00} & M_{01} \\ 
        M_{10} & M_{11}
    \end{pmatrix},
        \label{tenProd4}
\end{equation}
then $N \ot M = $
\begin{equation}
    \begin{pmatrix} 
        N_{00} 
        \begin{pmatrix} 
            M_{00} & M_{01} \\ 
            M_{10} & M_{11}
        \end{pmatrix},
        &
        N_{01} 
        \begin{pmatrix} 
            M_{00} & M_{01} \\ 
            M_{10} & M_{11}
        \end{pmatrix},
        \\
        N_{10} 
        \begin{pmatrix} 
            M_{00} & M_{01} \\ 
            M_{10} & M_{11}
        \end{pmatrix},
        &
        N_{11} 
        \begin{pmatrix} 
            M_{00} & M_{01} \\ 
            M_{10} & M_{11}
        \end{pmatrix},
    \end{pmatrix}
    \quad = 
    \begin{pmatrix} 
        N_{00} M_{00} & N_{00} M_{01} & N_{01} M_{00} & N_{01} M_{01}\\
        N_{00} M_{10} & N_{00} M_{11} & N_{01} M_{10} & N_{01} M_{11}\\
        N_{10} M_{00} & N_{10} M_{01} & N_{11} M_{00} & N_{11} M_{01}\\
        N_{10} M_{10} & N_{10} M_{11} & N_{11} M_{10} & N_{11} M_{11}\\
    \end{pmatrix}.
\end{equation}

\subsection{Measurement and Born's Rule}
\label{BornRule}

In quantum mechanics, {\em physical variables} or {\em observables} are
represented by Hermitian operators. As stated earlier, a {\em Hermitian} or
{\em self-adjoint} operator $O$ is one that equals its adjoint $O^{\dag}$. The
simplest Hermitian operator is a {\em projector}.  A projector $P$ is both
Hermitian, $P^{\dag} = P$, and idempotent, $P^2 = P$. The simplest projector
has rank 1 and can be written as $P = \dya{\psi}/\Tr(\dya{\psi})$ where
$\ket{\psi}$ is any ket. In general, any Hermitian operator $O$~(representing
some observable) has a {\em spectral decomposition} using which it can be
written as the sum of orthogonal projectors,
\begin{equation}
    O = \sum_i \lm_i P_i
    \label{obsDec}
\end{equation}
where $\lm_i$ are distinct real numbers representing distinct eigenvalues of
$O$, the projectors $\{P_i\}$ satisfy $P_i P_j = \dl_{ij}P_j$ ---that is, they are
orthogonal--- and $\sum_i P_i = I$--- that is, $\{P_i\}$ form a {\em projective
decomposition} of the identity. If an observable $O$ is measured on a system
with state $\ket{\phi}$, then one obtains its eigenvalue $\lm_i$ as a
measurement outcome.  According to Born's rule, the probability of observing the
value $\lm_i$ is
\begin{equation}
    p_i = \Tr(P_i \dya{\phi}).
    \label{bornRule}
\end{equation}
One simple observable is the identity $I$. Its decomposition of the
form~\eqref{obsDec} contains a single projector $I$ corresponding to the
eigenvalue $1$. If $I$ is measured on a system with state $\ket{\phi}$, then
one obtains its eigenvalue, $1$, as a measurement outcome with probability $1 =
\Tr(I \dya{\phi})$. Another simple observable in two dimensions is $B =
\dya{1}$. The decomposition~\eqref{obsDec} for $B$ takes the form
\begin{equation}
    B = 0 \cdot \dya{0} + 1 \cdot \dya{1}.
    \label{obsBDec}
\end{equation}
When the observable $B$ is measured on the state $\ket{\phi}$ in~\eqref{qbitCol},
one obtains two outcomes, $0$ with probability $|c_0|^2 = \Tr(\dya{0}
\dya{\phi})$ and $1$ with probability $|c_1|^2 = \Tr(\dya{1} \dya{\phi})$.
Notice, the normalization condition, $|c_0|^2 + |c_1|^2 = 1$, stated
below~\eqref{qbitCol} ensures that the probabilities sum to one. This type
of measurement, which results in the measurement of $B$, is called
a measurement in the computational basis.

\section{Quantum Information Theory: The Bare Minimum}
\label{SQIBare}

An example of classical information is learning the outcome of an unbiased coin
toss. This outcome takes values from a two-letter alphabet $\XC = \{H,T\}$,
where $H$ and $T$ represent heads and tails.  The probability that the coin toss
result $X$ takes a value $x \in \XC$ is $p(x):= \Pr(X = x)$, where $p(H) = p(T)
= 1/2$ for an unbiased coin. For a general biased coin, $p(H) = p, p(T) = 1- p$
and $0 \leq p \leq 1$. Learning the coin toss outcome provides classical
information because this learning removes uncertainty in the outcome. The
amount of uncertainty in the outcome of a biased coin with $p(H) = p$ is
captured by the {\em binary entropy},
\begin{equation}
    h(p): =  -p \log_2 p - (1-p) \log_2 (1-p),
    \label{binEntro}
\end{equation}
measured in bits, where $0 \log 0 := 0$. When $p=1/2$, $h(p) = 1$ bit, a result
that agrees with the usual intuition that learning the outcome of an unbiased
coin provides $1$ bit of information. In general, any random variable $X$
taking values $x$ from a finite alphabet $\XC$ with probability $p(x)$ has
{\em Shannon entropy},
\begin{equation}
    H(X) = -\sum_{x \in \XC} p(x) \log_2 p(x).
    \label{shanEntro}
\end{equation}
The entropy $H(X)$ quantifies the amount of uncertainty in the random variable
$X$. Operationally, it represents the ultimate limit for compressing symbols
$x$ appearing with probability $p(x)$~(see~\cite{CoverThomas01} for additional
discussion).  For two random variable $X$ and $Y$ with joint probability mass
function $p(x,y)$, the joint entropy,
\begin{equation}
    H(X,Y) = -\sum_{x,y} p(x,y) \log_2 p(x,y),
    \label{jointEntro}
\end{equation}
captures the amount of uncertainty in both $X$ and $Y$ taken together as
a single random variable. For a given outcome $x$ of $X$, the probability
of obtaining $y$ is $p(y|x) = p(x,y)/p(x)$. This probability mass has entropy,
\begin{equation}
    H(Y|X=x) = -\sum_{y} p(y|x) \log_2 p(y|x),
    \label{cndEntro}
\end{equation}
representing the uncertainty in $Y$ given $x$. The average value of the entropy
above,
\begin{equation}
    H(Y|X) = \sum_{x} p(x) H(Y|X=x),
    \label{condEntro}
\end{equation}
is called the {\em conditional entropy} of $Y$ given $X$. Subtracting this conditional
entropy from the entropy of $Y$ gives
\begin{equation}
    I(X;Y) = H(Y) - H(Y|X) 
    \label{mi}
\end{equation}
the {\em mutual information} between $Y$ and $X$. The mutual information is
unchanged when $X$ and $Y$ in~\eqref{mi} are interchanged. For two probability
mass functions $p(x)$ and $q(x)$, where $q(x) = 0$ only if $p(x) = 0$, one
defines the {\em relative entropy},
\begin{equation}
    D(p||q) = \sum_{x} p(x) \log_2 \big( p(x)/q(x) \big),
    \label{relEntro}
\end{equation}
a measure of how different $q(x)$ is from $p(x)$.  Operationally, $D(p||q)$
quantifies the penalty of compressing symbols from a distribution $p(x)$,
assuming it is $q(x)$(see~\cite{CoverThomas01} for additional discussion).

The notion of entropy can be generalized to the quantum world. However, prior
to performing such a generalization, we discuss the quantum analog of a
probability distribution, called a {\em density operator} or a {\em mixed
state}. Consider a collection of $N$ unmarked quantum systems, $n_i>0$ of which
are in the quantum state $\ket{\psi_i}$; then a quantum system chosen uniformly
at random from this collection is assigned a density operator,
\begin{equation}
    \rho = \sum_i p_i \dya{\psi_i},
    \label{denOp}
\end{equation}
where $p_i = n_i/N$. By construction, the density operator $\rho$ above is
Hermitian, positive semi-definite~($\rho \succeq 0$ or $\bra{\phi} \rho
\ket{\phi} \geq 0$ for all $\ket{\phi}$), and has unit trace~($\Tr(\rho) = 1$).
In general, any positive semi-definite operator with unit trace can be written
in the form~\eqref{denOp}. In this form, if every $\ket{\psi_i}$ is some
fixed state $\ket{\chi}$, then $\rho = \dya{\chi}$ represents a pure quantum
state $\ket{\chi}$ and $\rho^2 = \dya{\chi} . \dya{\chi} = \inp{\chi} \dya{\chi}
=\dya{\chi} = \rho$. 
More generally, $\rho$ represents a mixed state and can be written as
\begin{equation}
    \rho = \sum_i \lm_i \dya{m_i},
    \label{rhoDec}
\end{equation}
where eigenvalues $\lm_i$ are real, positive, and sum to one and
$\{\ket{m_i}\}$ are orthonormal kets. Using the form of $\rho$
in~\eqref{rhoDec}, one can easily obtain $\rho^2$ by replacing each $\lm_i$
with $\lm_i^2$. Notice $\rho^2 = \rho$ if each $\lm_i^2 = \lm_i$, which happens
only when $\lm_j = 1$ for some fixed $j$ and zero for all others--- that is,
$\rho = \dya{m_j}$--- represents a pure state.
Another natural context for using density operators is to describe sub-systems
of larger quantum systems. Suppose a quantum system $ab$, composed of two
systems $a$ and $b$ with spaces $\HC_a$ and $\HC_b$, respectively, is in
a pure state $\ket{\chi} \in \HC_{ab}$. As mentioned at the end of
Sec.~\ref{SQMBare}, this state need not be the product of two pure states, one
each on $\HC_a$ and $\HC_b$. The state of the $a$ and $b$ systems is
represented by mixed states with density operators 
\textcolor{black}{
\begin{equation}
    \rho_a = \Tr_b(\rho_{ab}), \quad \text{and} \quad \rho_b = \Tr_a (\rho_{ab}),
    \label{denOp2}
\end{equation}
respectively, where $\rho_{ab} = \dya{\chi}$}, $\Tr_a$ is the partial trace over $\HC_a$; i.e., $\Tr_a(A
\ot R) = R \Tr(A)$ where $\ot$ is the tensor product~(see discussion
containing~\eqref{tenProd3} and~\eqref{tenProd4}) and $\Tr_{b}$ is the partial
trace over $\HC_b$, defined similarly.

The simplest example of a density operator is a {\em qubit
density} operator. Such density operators can be written in the Bloch parametrization,
\begin{equation}
    \rho(\rB) = \frac{1}{2} (I + x X + y Y + z Z),
    \label{blochVec}
\end{equation}
where the real three-dimensional vector $\rB:= (x,y,z)$, called the Bloch vector, 
has magnitude $|\rB| = \sqrt{\rB.\rB}$, at most 1 and 
\begin{equation}
    X = \mat{0 & 1 \\ 1 & 0}, \quad 
    Y = \mat{0 & -i \\ i & 0}, \quad
    Z = \mat{1 & 0 \\ 0 & -1}
    \label{pauliMat}
\end{equation}
are the Pauli matrices, written in the standard basis $\{\ket{0}, \ket{1}\}$.
Using the Bloch parametrization~\eqref{blochVec}, any qubit density operator
can be represented by its Bloch vector in a unit sphere called the Bloch
sphere~(see Fig.~\ref{fig:blochVec}). \textcolor{black}{For instance the
density operator $\dya{i}$ has Bloch vector $\rB = (0,0,(-1)^i)$}
A Bloch vector of unit length, $\rB.\rB=1$, represents a pure state qubit
density operator, in other words, $\rho(\rB)^2 = \rho(\rB)$. \textcolor{black}{If
the length of the Bloch vector is less than one, then $\rho(\rB)$ is a mixed
state.}

\textcolor{black}{To help understand the Bloch sphere picture, we focus on the
density operator in~\eqref{blochVec}. This density operator can be written in
the form~\eqref{rhoDec}
\begin{equation}
    \rho(\rB) = \lm \dya{m_1} + (1-\lm) \dya{m_{\textcolor{black}{2}}},
    \label{blochVec2}
\end{equation}
where $\lm = (1 + |\rB|)/2$ and $1-\lm$ are the eigenvalues of $\rho(\rB)$.
Notice the eigenvalues are non-negative if and only if $|\rB| \leq 1$.  The
eigenvectors $\ket{m_1}$ and $\ket{m_2}$, corresponding to eigenvalues $\lm$
and $1-\lm$, respectively, are normalized, $\inpd{m_1}{m_1} = \inpd{m_2}{m_2} =
1$, and orthogonal to each other, $\inpd{m_1}{m_2} = 0$, i.e., they represent
orthogonal pure states. While one can write explicit expressions for these pure
states, it is more useful to focus on the projectors $\dya{m_1}$ and
$\dya{m_2}$ onto these pure states~(see discussion above \eqref{rhoDec} for
additional discussion on projectors). These projectors represent density
operators and it is instructive to use the Bloch
parametrization~\eqref{blochVec} to represent them
\begin{equation}
    \dya{m_1} = \rho(\rB_1), \quad \text{and} \quad \dya{m_2} = \rho(\rB_2),
    \label{eq:m12}
\end{equation}
where $\rB_1 := \rB/|\rB|$ and $\rB_2 := -\rB/|\rB|$ are unit vectors. Using
the above equation in~\eqref{blochVec2}, we find $\rho(\rB)$ is the convex
combination of two pure state density operators,
\begin{equation}
    \rho(\rB) = \lm \rho(\rB_1) + (1-\lm) \rho(\rB_2).
    \label{blochVec3}
\end{equation}
In addition, one finds that the Bloch vector $\rB$ is a convex combination,
$\rB = \lm \rB_1 + (1- \lm) \rB_2$, of two unit vectors $\rB_1$ and $\rB_2$.
The Bloch sphere picture~(see Fig.~\ref{fig:blochVec}) provides a simple way to
visualize all three density operators $\rho(\rB), \rho(\rB_1)$ and
$\rho(\rB_2)$.}

More generally, two systems $a$ and $b$, each with density operators $\rho_a$ and
$\rho_b$, respectively, are assigned a joint density operator 
\begin{equation}
    \rho_{ab} = \rho_a \ot \rho_b,
    \label{tenProd5}
\end{equation}
where $\ot$ represents tensor product~(see discussion
containing~\eqref{tenProd3} and~\eqref{tenProd4}).

\begin{figure}[h]
    \centering
    \begin{tikzpicture}[x=1cm,y=1cm,z=0.6cm,>=stealth, scale=2]
    \draw[->] (xyz cs:x=0) -- (xyz cs:x= 1.2) node[above] {$y$};
    \draw[<->] (xyz cs:y=-1.1) -- (xyz cs:y= 1.1) node[right] {$z$};
    \draw[->] (xyz cs:z=0) -- (xyz cs:z= -1.4) node[left] {$x$};

    \coordinate (O) at (0, 0) {};
    \draw[line width=.05mm] (O) circle(1);
    \draw [dashed, line width=.05mm] (O) ellipse(1 and .25);
    \node[] at (0,1.125) (Ot) {};
    \node[] at (0,-1.125) (Ob) {};
    \node[] at (0,1.25) (Otk) {\normalsize $[0]$};
    \node[] at (0,-1.25) (Obk) {\normalsize $[1]$};

    \draw[color=black, thick,->] (0,0,0) -- (0.1,0.3,0.2) node[above] {$\rB$};
    \draw[dotted,color=black, thin,->] (0,0,0) -- (0.21,0.63,0.42) node[above] {$\rB_1$};
    \draw[dotted,color=black, thin,->] (0,0,0) -- (-0.21,-0.63,-0.42) node[below] {$\rB_2$};

    \end{tikzpicture}
    \caption{Bloch sphere in xyz Cartesian coordinates: The Bloch vector $\rB$
    represents a qubit density operator $\rho(\rB)$~\eqref{blochVec}.
    \textcolor{black}{The other Bloch vectors, $\rB_{1}$ and $\rB_{2}$~(defined
    below~\eqref{eq:m12}), have unit length.  These vectors $\rB_1$ and $\rB_2$
    are antipodes of each other and represent projectors, $\dya{m_1}$ and
    $\dya{m_2}$, respectively. Such rank-1 projectors can be written using a
    square bracket notation; for instance, $[0]:= \dya{0}$.}}
    \label{fig:blochVec}
\end{figure}

The quantum analog of the Shannon entropy~\eqref{shanEntro} is the von-Neumann
entropy of a density operator $\rho$~\eqref{rhoDec},
\begin{equation}
    S(\rho) := - \Tr (\rho \log \rho) = -\sum_i \lm_i \log \lm_i.
    \label{entVon}
\end{equation}
\textcolor{black}{The inequality above can be derived using the
eigen-decomposition~\eqref{rhoDec}, where the eigenvalues $\lm_i$ of $\rho$ are
strictly positive. Using this decomposition, one obtains $\log \rho := \sum_i
\log \lm_i \dya{m_i}$ by simply replacing the eigenvalues $\lm_i$ with their
logarithm.  Multiplying $\log \rho$ with $\rho$ gives $\rho \log \rho = \sum_i
\lm_i \log \lm_i \dya{m_i}$. The trace of this product, multiplied with minus,
one gives the right side of the equality~\eqref{entVon}}
For the qubit density operator in~\eqref{blochVec} the von-Neumann entropy can
be easily computed using~\eqref{blochVec2} as 
\begin{equation}
    S\big( \rho (\rB) \big) = h(\lm_{+}) := -\big( \lm_{+} \log \lm_{+} + (1-\lm_{+}) \log (1-\lm_{+}) \big),
    \label{shEnt}
\end{equation}
where we use $\lm_{-} = 1 - \lm_{+}$.
Much like the Shannon entropy, the von-Neumann entropy quantifies the amount of
uncertainty in the quantum state $\rho$.  In addition, the von-Neumann entropy
plays a fundamental role in a vast variety of information processing tasks.
For instance, it represents the ultimate rate for compressing quantum
states~\cite{Schumacher95}.

In practice, quantum systems are susceptible to noise. Prior to describing
quantum noise, let us consider classical noise. Classical noise is often
modelled by a channel $N:\XC \to \YC$, which maps some input symbol $x \in
\XC$ to an output symbol $y \in \YC$ with probability $p(y|x)$~(see
Sec.~\ref{P5Cap} for more details).
If a channel's input symbol $\XC$ arrives with some probability $p(x)$ then the
channel maps this input distribution to an output distribution $p(y) = \sum_x
p(y|x)p(x)$ over the channel outputs $\YC$. 

\textcolor{black}{Consider a simple example of a classical erasure channel $E$
with erasure probability $p$. The channel's input alphabet $\XC = \{0,1\}$ and
output alphabet $\YC = \{0,1,e\}$. With probability $p$, the channel erases the
input $x$ by mapping it to an output $y=e$, otherwise with probability $1-p$
the input is sent perfectly, i.e., the output $y = x$.  This erasure channel's
conditional probability distribution $p(y|x)$ is given as follows: $p(e|0) =
p(e|1) = p, p(0|0) = p(1|1) = (1-p)$, and $p(1|0) = p(0|1) = 0$. Using this
conditional probability distribution, one can find the output distribution
\begin{equation}
    p(y) = (1-p) \; p(x)\dl_{y,x} + p \dl_{y,e}.
    \label{eq:classicalErasure}
\end{equation}
If $x = i$, where $i \in \{ 0,1 \}$, then $p(y=i) = 1-p, p(y=1-i)
=0$, and $p(y = e) = p$.}

The quantum analog of a channel, called a quantum channel, describes quantum
noise~\cite{SudarshanMathewsEA61}. Like its classical counterpart, a quantum
channel acting on the quantum analog of a probability distribution, a density
operator, maps it to another density operator. 
In addition, a quantum channel acting on one part of a bi-partite density
operator maps the bi-partite density operator to a valid density operator.
Mathematically, a quantum channel is a completely positive trace preserving~(CPTP)
map~(see discussion below~\eqref{Qchan}).
One simple example of a quantum channel is an {\em erasure channel}.  Let $a$
be a $d_a$-dimensional quantum system with space $\HC_a = \HC$ and let $b$ be a
$d_b=(d_a+1)$-dimensional quantum system with space $\HC_b = \HC \oplus \HC'$,
where $\HC'$ is spanned by multiples of a single ket $\ket{e}$. Then an erasure
channel with erasure probability $p$, $\EC_{p}: \LC(\HC_a) \to
\LC(\HC_b)$, is given by 
\begin{equation}
    \EC_{p}(\rho) = (1-p) \rho + p \Tr(\rho) \dya{e},
    \label{eraChan}
\end{equation}
where the channel input $\rho$ is sent perfectly with probability $1-p$;
otherwise, the input is erased with probability $p$ and mapped to a fixed pure
state $\dya{e}$ orthogonal to $\rho$. \textcolor{black}{The above equation is akin
to~\eqref{eq:classicalErasure} where the output probability
distribution $p(y)$ was expressed in terms of the input probability distribution $p(x)$ for
the classical erasure channel $E$. The connection with classical erasure can
be made even tighter: a classical erasure channel $E$ can emerge from a
quantum erasure channel $\EC_p$ in the following sense. Suppose a classical
input symbol $x = i$, $i \in \{0,1\}$ is mapped to a quantum state $[i]$, then
sent via $\EC_p$, and finally measured using projectors $\{P_0 = [0],P_1 =
[1],P_e = [e]\}$ corresponding to measurements outcomes $y \in \{0,1,e\}$
respectively.  Then, using the Born rule~\eqref{bornRule}, the expression for
$\EC_p$~\eqref{eraChan}, and the definitions of the projectors $\{P_j\}$, one
finds that $p(y|x)$, the probability that the measurement outcome is $y$ given the
input symbol is $x$, can be simply written as $p(y=j|x=i) = \Tr( \EC_{p}([i])
[j])$. This conditional probability is exactly the same as the one for the
erasure channel $E$ above~\eqref{eq:classicalErasure}.}

\begin{figure}[h]
    \centering
    \begin{tikzpicture}[scale=2]

    \path (0,-1) edge[line width=.1mm, <->] node[minimum size=.1mm, fill=white, anchor=center, pos=0.5, font=\normalsize] {$z$ axis}(0,1);    

    \coordinate (O) at (0, 0) {};
    \draw[line width=.05mm] (O) circle(1);
    \draw [dashed, line width=.05mm] (O) ellipse(1 and .25);
    \node[] at (0,1.125) (Ot) {};
    \node[] at (0,-1.125) (Ob) {};
    \node[] at (0,1.2) (Otk) {\normalsize $[0]$};
    \node[] at (0,-1.2) (Obk) {\normalsize $[1]$};
        
    \coordinate (Ob) at (0, 0) {};
    \draw [line width=.1mm, red] (Ob) circle(.7);
    \draw [line width=.1mm, red, dashed] (Ob) ellipse(.7 and .175);
    \end{tikzpicture}
    \caption{Qubit depolarizing channel: The qubit Bloch sphere~(in black) is
    transformed into to an origin centered sphere with a shorter radius~(in
    red) under the action of a qubit depolarizing channel~\eqref{depol}.  }
    \label{fig:depolCh}
\end{figure}
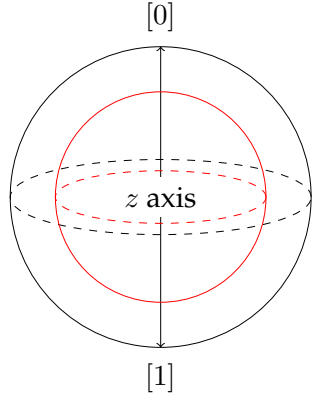

Yet another example of a quantum channel is a {\em qubit depolarizing} channel
$\Dl: \LC(\HC_a) \to \LC(\HC_b)$ where $d_a = d_b = 2$,
\begin{equation}
    \Dl(\rho) = \lm \rho + \frac{(1- \lm)}{2} \Tr (\rho) I,
    \label{depol}
\end{equation}
and $-1/3 \leq \lm \leq 1$.
\textcolor{black}{There is no obvious analogy between this quantum depolarizing
channel and a classical channel, however the Bloch sphere picture, discussed
below~\eqref{blochVec}, provides a helpful way to visualize the action of the
depolarizing channel.}
The depolarizing channel takes its qubit input $\rho$, with Bloch vector $\rB$,
to a qubit output $\Dl(\rho)$ with Bloch vector $\lm \rB$. \textcolor{black}{The
effect of the channel is to scale the Bloch sphere and make its length
smaller}~(see Fig.~\ref{fig:depolCh} for a graphical representation). For
values of $\lm \geq 0$, this channel is often interpreted as sending its input
perfectly with probability $\lm$ or replacing the input with the maximally mixed
state $I/2$. There are a large variety of well-studied quantum
channels~\cite{Holevo12, Wilde17, Watrous18}. Some common ones include the
qubit dephasing channel and the (generalized)~qubit amplitude damping
channel~\cite{WolfPerezGarcia07, KhatriSharmaEA20}. 

To describe the action of two separate quantum channels $\BC$ and $\BC'$ acting
on systems $a$ and $a'$, respectively, one uses tensor products in a
manner similar to those used for describing pure states and mixed state on two
systems~(see discussion containing~\eqref{tenProd1} and~\eqref{tenProd5}).
Let systems $a$ and $a'$ be acted upon by quantum channels $\BC : \LC(\HC_a)
\to \LC(\HC_{b})$ and $\BC' : \LC(\HC_{a'}) \to \LC(\HC_{b'})$,
respectively, then the channel acting on the joint $aa'$ system is the tensor
product channel $\BC \ot \BC'$. The tensor product channel is linear, and if $\rho_a$ and $\rho_{a'}$ are density
operators of $a$ and $a'$, respectively, then 
\begin{equation}
    \BC \ot \BC' (\rho_{a} \ot \rho_{a'}) = \BC(\rho_a) \ot \BC'(\rho_{a'}).
    \label{tenProd6}
\end{equation}

\noindent {\bf Remark.} When discussing the transmission of classical
information across a quantum channel, one arrives at {\em induced classical
channels}~(see Ch.20 in~\cite{Wilde17}). These classical channels $N:\XC
\to \YC$ arise out of quantum ones and they model the effective classical
noise experienced by classical information encoded and decoded into quantum
states passing through a quantum channel $\BC$.  The {\em capacity} $C(N)$ of
any such induced channel $N$ is obtained by maximizing the mutual
information~\eqref{mi} between the channel output $Y$ and input $X$ over all
possible input distributions $p(x)$. This capacity is bounded from above by the
{\em Holevo capacity} of the channel $\BC$, which represents the ultimate rate
at which classical information can be sent across a quantum channel without
entanglement at the input~\cite{HolevoGiovannetti12}. This is not the subject
of any of the five SDP problems presented here.

\vspace{0.1in}

In general, any quantum channel $\BC : \LC(\HC_a) \to \LC(\HC_b)$, where
$\HC_a$ and $\HC_b$ have possibly unequal dimensions, can be written using a
{\em Kraus decomposition}~\cite{Kraus83}
\begin{equation}
    \BC(A) = \sum_i K_i A K_i^{\dag},
    \label{Qchan}
\end{equation}
where $A \in \LC(\HC_a)$, and $K_i: \HC_a \to \HC_b$ are Kraus
operators that satisfy the relation
\begin{equation}
    \sum_i K_i^{\dag} K_i = I_a.
    \label{krausOp}
\end{equation}
For an erasure channel of the form~\eqref{eraChan} acting on qubits inputs, the
Kraus operators can be written as matrices 
\begin{equation}
    K_1 = 
    \sqrt{1 - p}
    \begin{pmatrix} 
        1 & 0 \\
        0 & 1 \\
        0 & 0
    \end{pmatrix},
    \quad
    K_2 = 
    \sqrt{p}
    \begin{pmatrix} 
        0 & 0 \\
        0 & 0 \\
        1 & 0
    \end{pmatrix},
    \quad
    K_3 = 
    \sqrt{p}
    \begin{pmatrix} 
        0 & 0 \\
        0 & 0 \\
        0 & 1 
    \end{pmatrix},
\end{equation}
using the standard basis $\{ \ket{0}, \ket{1} \}$ at the input and the basis
$\{\ket{0}, \ket{1}, \ket{e}\}$ at the output. A simple calculation shows that
the matrices above satisfy~\eqref{krausOp} with $I_a$ as the $2 \times 2$
identity matrix $I_2$. The qubit depolarizing channel $\Delta$~\eqref{depol}
can also be written in the Kraus form~\eqref{Qchan} where
\begin{equation}
    K_1 = \sqrt{1-p} I_2, \quad 
    K_2 = \sqrt{p/3} X, \quad
    K_3 = \sqrt{p/3} Y, \quad
    K_4 = \sqrt{p/3} Z,
    \label{krausOpDepol}
\end{equation}
and $p = 3(1-\lm)/4$. From this Kraus form, one may view $\Delta(\rho)$ as a channel that 
applies each of the Pauli errors $X,Y,$ and $Z$ with
equal probability $p/3$ and applies the identity map with probability $1-p$.
Using standard matrix multiplication, or the property $X^\dag X = Y^\dag Y =
Z^\dag Z = I_2$, one can check that the operators in~\eqref{krausOpDepol} also
satisfy the equality in~\eqref{krausOp}.
This equality~\eqref{krausOp} ensures that $\BC$ is trace preserving; that is,
$\Tr(\BC(A)) = \Tr(A)$ for any operator $A \in \LC(\HC_a)$.
Together,~\eqref{krausOp} and the Kraus decomposition~\eqref{Qchan} ensure that
$\BC$ is a completely positive trace preserving map. While a positive trace
preserving map is one that maps positive semi-definite operators to positive
semi-definite operators of the same trace, a CPTP map satisfies a stronger
condition: for all positive semi-definite operators $\Lm_{ar}$ and any finite-dimensional space $\HC_{r}$ with dimension $d_r$, the operator
\begin{equation}
    \Gm_{br} = (\BC \ot \IC ) \Lm_{ar},
    \label{ChoiJPre}
\end{equation}
is positive semi-definite and has the same trace as $\Lm_{ar}$, where
$\IC$ is the identity channel taking $\LC(\HC_r)$ to itself and $\BC \ot \IC$
is a tensor product of two channels $\BC$ and $\IC$~(see discussion
containing~\eqref{tenProd6}) . Turns out this stronger condition is satisfied
if and only if at $\HC_r = \HC_a$ and $\Lm_{aa} = \dya{\gm}$, where $\ket{\gm}
= \sum_i \ket{i}_a \ot \ket{i}_a$ is an {\em unnormalized} maximally entangled
state across $\HC_a \ot \HC_a$, the operator $\Gm_{ba}$ is positive
semi-definite and its partial trace over $b$ is the identity on $\HC_a$, i.e.,
$\Tr_{b}(\Gm_{br}) = I_a$~\cite{Choi75, Jamiokowski72}. For this reason, an
operator \textcolor{black}{mapping $\HC_{ba}:= \HC_b \ot \HC_a$ to itself},
\begin{equation}
    \JC_{ba}(\BC) := (\BC \ot \IC) \dya{\gm},
    \label{ChoiJ}
\end{equation}
is sometimes called the Choi-Jamio\l{}kowski representation of the channel
$\BC$.  Given two quantum channels $\BC: \LC(\HC_a) \to \LC(\HC_b)$ and
$\BC': \LC(\HC_a) \to \LC(\HC_b)$, their linear combination, $\SC = c_0 \BC
+ c_1 \BC'$, $c_0,c_0 \in \mathbb{C}$, is a linear map from $\LC(\HC_a)$ to
$\LC(\HC_b)$, however this map need not be a quantum channel.  In general, a
linear map $\SC: \LC(\HC_a) \to \LC(\HC_b)$ from operators on $\HC_a$ to
operators on $\HC_b$ is called a {\em superoperator}. This superoperator
represents a CPTP map if and only if its Choi-Jamio\l{}kowski representation,
$\JC_{ba}(\SC)$, is positive semi-definite and $\Tr_{b}(\JC_{ba}(\SC)) = I_a$.

The Kraus form~\eqref{Qchan} ensures that a map is CPTP, but it can be also
used to interpret properties of a quantum channel. For instance, if each Kraus
operator in~\eqref{Qchan} has rank 1, then $\BC$ becomes an {\em entanglement
breaking} channel~\cite{HorodeckiShorEA03}. Such channels have the property
that any entangled channel input $\rho_{ra}$ is mapped to an unentangled
output~\eqref{ChoiJ}~(see Sec.~\ref{P4EntSep} for additional discussion about
entanglement).
One simple example of an entanglement breaking channel is a qubit
channel $\EC$ with two Kraus operators:
\begin{equation}
    K_1 = \dya{0} = \begin{pmatrix} 1 & 0 \\ 0 & 0\end{pmatrix} \quad
        \text{and} \quad
    K_2 = \dya{1} = \begin{pmatrix} 0 & 0 \\ 0 & 1\end{pmatrix}.
        \label{krausEBT}
\end{equation}
Each operator above has rank $1$. To express $\EC$ using these operators
in~\eqref{krausEBT}, one uses and simplifies an equation of the
form~\eqref{Qchan} to obtain
\begin{equation}
    \EC(\rho) =  \dya{0} \Tr(\rho \dya{0}) + \dya{1} \Tr(\rho \dya{1}).
    \label{EBTChan}
\end{equation}

\section{Problem 1: Quantum State Discrimination}
\label{P1StDis}

The concept of projective measurements, discussed in Sec.~\ref{BornRule}, and
quantum channels, discussed at the end of Sec.~\ref{SQIBare}, can be combined
to obtain a more general measurement scheme described mathematically as a
positive operator value measure~(POVM). Consider a quantum channel $\BC :
\LC(\HC_a) \to \LC(\HC_b) \ot \LC(\HC_e)$, of the form
\begin{equation}
    \BC(\rho) = \sum_i K_i \rho K_i^{\dag} \ot \dya{i},
    \label{chanMes}
\end{equation}
where $\ot$ represents tensor product~(see discussion containing~\eqref{tenProd3}
and~\eqref{tenProd4}). 
Using Born's rule, a projective measurement on one half of the channel output
$\HC_e$ in the computational basis of $\HC_e$ results in an outcome $i$ with
probability
\begin{equation}
    p_i = \Tr\big( \BC(\rho) (I_b \ot \dya{i}) \big).
\end{equation}
Using standard linear algebra along with the definition $E_i := K_i^{\dag} K_i$,
we obtain
\begin{equation}
    p_i = \Tr( \rho E_i).
    \label{bornGln}
\end{equation}
The collection of operators $\{E_i\}$ are called a POVM. These operators are
positive semi-definite and sum to the identity $I_a$; that is, 
\begin{equation}
    E_j = E_j^{\dag} \succeq 0, \quad \text{and} \quad
    \sum_i E_i = I_a.
    \label{POVM}
\end{equation}
Any general measurement on a quantum system $a$ can be described using a POVM
$\{E_i\}$.  Associated with each $E_i$ is a measurement outcome $i$ which
occurs with probability~\eqref{bornGln}.

A general setup for the quantum state discrimination problem can be obtained
as follows. Suppose a random variable $X$ takes one of $n$ values $i$ with
probability $p_i$.  When $X=i$, a $d$-dimensional quantum state $\sg_i$ is
prepared. The key task in quantum state discrimination is to measure the
prepared state and predict $i$ with high probability. If the random variable
$Y$ predicts $X$, then we wish to maximize the success probability: 
\begin{equation}
    p_s:=\sum_i p_i \Pr(Y = i| X = i). 
    \label{pSuccess}
\end{equation}
Suppose $\{E_j\}$ is a POVM that describes the measurement; then the
conditional probability 
\begin{equation}
    \Pr(Y=j|X=i) = \Tr(E_j \sg_i). 
    \label{cdtProb}
\end{equation}
Using the above equation, the maximum success probability~\eqref{pSuccess} over
all POVMs is obtained as the optimum value $p_s^*$ of the semi-definite 
program,
\begin{align}
    \label{SDPStateDisc}
    \begin{aligned}
        \text{maximize} \; &  \sum_i p_i \Tr(E_i \sg_i) & \\ 
        \text{subject to} \; & E_j \succeq 0,& \forall 1 \leq j \leq n, \\ 
        \text{and} \; & \sum_{j=1}^{n} E_j =I. &
    \end{aligned}
\end{align}

Consider a simple case when $n = d = 2$, $p_1 = q, p_2 = 1-q$,
\begin{equation}
    \sg_1 = \begin{pmatrix}
        1 & 0 \\
        0 & 0
    \end{pmatrix},
    \quad \text{and} \quad
    \sg_2 = \begin{pmatrix}
        0 & 0 \\
        0 & 1
    \end{pmatrix}.
    \label{eq:sgs}
\end{equation}
In this case, the SDP in~\eqref{SDPStateDisc} admits an algebraic solution, $E_1
= \sg_1, E_2 = I - E_1$ and $p_s = 1$, which is independent of $p$. Such
algebraic solutions exist for any $d \geq 2$ and $2 \leq n \leq d$ when 
\begin{equation}
    \Tr(\sg_i \sg_j) = \dl_{ij}
    \label{eq:piarOr}
\end{equation}
for all $1 \leq i,j \leq n$; that is,  when $\sg_i$ are pairwise
orthogonal to one another. In such cases $E_i = \sg_i$ for all $1 \leq i \neq
n-1$ and $E_n = I - \sum_i E_i$ is a solution to~\eqref{SDPStateDisc} with
optimum value $p_s^* = 1$, independent of $p_i$. 
An interesting case where the
SDP in~\eqref{SDPStateDisc} can be solved algebraically is $n = 2$ and
arbitrary $d$. In this case, the SDP in~\eqref{SDPStateDisc} reduces to
\begin{align}
    \label{SDPStateDisc2}
    \begin{aligned}
        \text{maximize} \; & \frac{1}{2}( 1 + \Tr\big(F(p_1\sg_1 - p_2\sg_2)\big) \\ 
        \text{subject to} \; & -I \preceq F \preceq I,
    \end{aligned}
\end{align}
where we have introduced the operator $F := E_1 - E_2$. The above SDP has an
optimum value
\begin{equation}
    p^* = \frac{1}{2}(1 + ||p_1\sg_1 - p_2\sg_2||_1),
    \label{eq:pr1Alg}
\end{equation}
where $||A||_1 = \Tr \sqrt{A^{\dag}A}$ is the operator $1$-norm,
\textcolor{black}{also called the nuclear norm}, and $\sqrt{A^{\dag}A}$, square
root of a positive semi-definite matrix, is an operator obtained by replacing
the eigenvalues of $A^{\dag}A$ with their square root. The optimum value is
often called the Helstrom bound~\cite{Helstrom69}.  In general, the SDP
in~\eqref{SDPStateDisc} cannot be solved analytically; however, one can use
numerical SDP solvers. For using such numerical solvers, we reformulate the SDP
in~\eqref{SDPStateDisc} as follows
\begin{align}
    \label{SDPStateDisc3}
    \begin{aligned}
        \text{maximize} \; &  \sum_{i=1}^{n-1} p_i \Tr(E_i \sg_i) +  (1-\sum_{i=1}^{n-1} p_i) \Tr\big( (I - \sum_{j=1}^{n-1} E_j) \sg_{n}\big) & \\
        \text{subject to} 
        \; & I - \sum_{j=1}^{n-1} E_j   \succeq 0, \\
        \text{and} 
        \;&  E_j  \succeq 0, & 
    \end{aligned}
\end{align}
where $1 \leq j \leq n-1$.  Numerical solutions to the SDP above can be
obtained using solvers in open-source packages. 
For instance, one may use a Python interface, PICOS~\cite{SagnolStahlberg21}, and an
open-source solver, CVXOPT~\cite{AndersenDahlEA21}.  Using these numerical
tools we formulate the SDP above with $n = d = 2$, $p \in [0,1]$ chosen
randomly, and $\sg_i$ defined in~\eqref{eq:sgs}.  We find almost perfect
agreement between the numerically obtained objective value $p_s^n$ and the
algebraic value $p^*s=1$ stated below~\eqref{eq:piarOr}. In other examples
with fixed $n=2, d=6$, and randomly chosen $p \in [0,1]$, $\sg_1$, and $\sg_2$,
we find good agreement, $|p_s^n - p_s*| \simeq O(10^{-9})$, between the
numerical value $p_s^n$ and the true value, $p_s*$, computed
using~\eqref{eq:pr1Alg}.
These and additional examples are available along with this chapter~(see
Notebook~1 in~\cite{NoName21a}).

Discrimination of quantum states, and quantum hypothesis testing, is a
vast~\cite{Helstrom69, Kholevo79, Kargin05, AudenaertCalsamigliaEA07,
Hayashi17, NussbaumSzkoa09, Parthasarathy01} and active sub-field of quantum
information science. Here, we have touched the surface of this field by
introducing certain special cases.  For solving these special cases, we
illustrate the use of open-source numerical tools. The interested reader may
find a variety of other resources and open problems in these
reviews~\cite{Chefles00a, BarnettCroke09, BaeKwek15} and references therein.

\section{Problem 2: Quantum State Fidelity}
\label{P2StFid}

Classical objects of different types are perfectly distinguishable. On the
other hand, quantum objects in two different quantum states are not always
perfectly distinguishable.  This motivates a basic question, given two
different states of a quantum system: how similar are these states to each
other? One measure of similarity between quantum states is {\em fidelity}.
Consider pure states $\ket{\psi}$ and $\ket{\phi}$; these are simply unit
vectors in some space $\HC_a$.  The magnitude of the overlap between these
vectors,
\begin{equation}
    F(\psi, \phi) = |\inpd{\psi}{\phi}|,
    \label{FIDPure}
\end{equation}
is defined as the fidelity between the pure states $\ket{\psi}$ and
$\ket{\phi}$. When the pure states  $\ket{\psi}$ and $\ket{\phi}$ are the same,
their fidelity $F(\psi, \phi)$ is one; when $\ket{\psi}$ and $\ket{\phi}$ are
orthogonal, $F(\psi, \phi) =0$, and in general, $0 \leq F(\psi, \phi) \leq
1$. 

While fidelity between pure states is straightforward to define, quantum
systems cannot always be described by pure states. In general, quantum systems
are described by mixed states. However, a quantum system $a$ in some mixed
state $\rho$ can always be viewed as a sub-system of two systems $a$ and $r$ in
some pure state $\ket{\psi}$. More precisely, let $\HC_a$ and $\HC_r$ describe the 
spaces of $a$ and $r$, respectively, and then any $\rho$ describing $a$ can be
obtained as
\begin{equation}
    \rho = \Tr_r(\dya{\psi}),
\end{equation}
where $\ket{\psi} \in \HC_{ar}$.  The state $\ket{\psi}$ above is called a
purification of $\rho$ and $r$, the purifying system. If $\rho$ has a spectral
decomposition $\rho = \sum_i \lm_i \dya{e_i}$ then its purification has the
form $\ket{\psi} = \sum_i \sqrt{\lm_i} \ket{e_i} \ot U \ket{e_i}$ where $U$ is
any unitary matrix on $\HC_r$ and $\ot$ represents tensor product~(see discussion
containing~\eqref{tenProd1} and~\eqref{tenProd2}). 
Clearly, every choice of $U$ gives a different purification of the same state
$\rho$ and two different purifications are related by a unitary on the
purifying system $\HC_r$. To define the fidelity between two, possibly mixed
states $\rho$ and $\sigma$ of a system $a$, we can consider the maximal
fidelity between their purifications,
\begin{equation}
    F(\rho,\sg) = \max_{\psi,\phi} |\inpd{\psi}{\phi}|,
    \label{FIDMixed1}
\end{equation}
where $\ket{\psi}$ and $\ket{\phi}$ purify $\rho$ and $\sg$ respectively, using
the same purifying system $r$. Uhlmann's theorem~\cite{Uhlmann76}~(see Th.9.2.1
in~\cite{Wilde17} for a short proof) shows that the fidelity defined above
reduces to the simple form
\begin{equation}
    F(\rho,\sg) = ||\sqrt{\rho}\sqrt{\sg}||_1.
    \label{FIDMixed2}
\end{equation}
When $\rho$ and $\sg$ are pure states $\dya{\psi}$ and $\dya{\phi}$, respectively,
the fidelity expression \eqref{FIDMixed2} simply reduces to~\eqref{FIDPure}.
When $\rho$ is a mixed state but
$\sg = \dya{\psi}$ is a pure state then the fidelity~\eqref{FIDMixed2} is simply
\begin{equation}
    F(\rho,\sg) = \bra{\psi} \rho \ket{\psi}.
    \label{FIDMixedPure}
\end{equation}
Suppose $\rho$ and $\sg$ are mixed states that are diagonal in the same basis
$\ket{x}$. It is convenient to write these states as follow
\begin{equation}
    \rho = \sum_x p(x) \dya{x}, \quad \text{and} \quad
    \sg = \sum_x q(x) \dya{x}
\end{equation}
where $p(x)$ and $q(x)$ are probability distributions. The fidelity between
these diagonal operators above 
\begin{equation}
    F(\rho,\sg) = \sum_x \sqrt{p(x)} \sqrt{q(x)},
    \label{FIDGln}
\end{equation}
is simply the Bhattacharya overlap~\cite{Bhattacharya43} between the classical
probability distributions $p(x)$ and $q(x)$. 

Fidelity $F(\rho,\sg)$~\eqref{FIDMixed2} is given by the optimum value of
these primal and dual semi-definite programs~\cite{Killoran12, Watrous12},

\begin{minipage}[t]{0.5\textwidth}
    \begin{equation}
        \text{Primal:}
    \end{equation}
    \begin{align*}
        \begin{aligned}
            \text{maximize} \; & \frac{1}{2}\Tr(\Lambda + \Lambda^{\dag}) \\
            \text{subject to} \; & 
            \begin{pmatrix}
            \rho & \Lambda \\ 
            \Lambda^{\dag} & \sg
            \end{pmatrix}
            \succeq 0, & \\ 
            \text{and}\;& \Lambda \in \LC(\HC_a);
        \end{aligned}
    \end{align*}
\end{minipage}
\begin{minipage}[t]{0.5\textwidth}
    \begin{equation}
        \text{Dual:}
        \label{SDPFid}
    \end{equation}
    \begin{align*}
        \begin{aligned}
        \text{minimize} \; &\frac{1}{2}\big(\Tr(\rho Y ) + \Tr(\sg Z)\big) &\\
         \text{subject to} \; &
         \begin{pmatrix}
         Y & -I \\ 
         -I & Z
        \end{pmatrix}
            \succeq 0. & \\
        \end{aligned}
    \end{align*}
\end{minipage}
The primal and dual SDPs can be solved numerically. When $\rho$ and $\sg$ are
randomly chosen 4-dimensional pure states we find optimum values $F_p$ and
$F_d$, for the primal and dual SDP objectives respectively. These are in good
agreement with each other, and with $F$ computed using~\eqref{FIDPure}. In
particular, the maximum absolute difference between any pair of these three
values is $O(10^{-9})$.
This absolute difference remains typically small, $O(10^{-5})$, when $\rho$ is
a random mixed state and $\sg$ is a random pure state, each 3-dimensional. To
compute this difference numerically we solve the primal and dual SDP above and
find $F$ using~\eqref{FIDMixedPure}. A similar computation, using 10-dimensional
mixed states $\rho$ and $\sg$ chosen randomly and $F$ computed 
using~\eqref{FIDGln}, shows good numerical agreement. Typically
the maximum pairwise difference between all three values $F_p, F_d$, and $F$, is $O(10^{-7})$.
A short tutorial helping perform these computations is available along with
this chapter~(see Notebook~2 in~\cite{NoName21a}).

While the fidelity function~\eqref{FIDMixed2} can be computed using standard
numerical algebra libraries, the SDP formulation~\eqref{SDPFid} for computing
fidelity has additional utilities. One utility is the use of the
formulation~\eqref{SDPFid} to show that a variety of generalizations of the
fidelity function~\cite{ZhangChenEA15,YuanFung17, GutoskiRosmanisEA18,
KatariyaWilde21} can also be computed efficiently via an SDP. Some of these
generalizations play a useful rule in security analysis of quantum protocols.
Another utility of the SDP formulation is to the quantum channel discrimination
problem discussed next.

\section{Problem 3: Quantum Channel Discrimination}
\label{P3ChDis}

Consider a protocol with two parties, Alice and Bob, where Alice prepares a
quantum state $\rho_a$ and hands it to Bob. Upon receiving $\rho_a$, Bob
generates a random variable $X$, which takes the value $1$ with probability $t$
and $2$ with probability $1 - t$.  When $X=i$, Bob applies the channel $\BC_i:
\hat \HC_a \to \hat \HC_b$ and obtains a state $\BC_i(\rho_a)$.  This new
state is returned to Alice, whose task is to measure it and correctly predict
$i$.  Alice knows a description of each fixed channel $\BC_i$ and controls the
state $\rho_a$, but is unaware of the random value $i$. By varying $\rho_a$ the
maximum probability with which Alice can correctly predict $i$ by measuring
$\BC_i(\rho_a)$ is
\begin{equation}
    q^* = \frac{1}{2} (1 + \max_{\rho_a}||t \BC_1(\rho_a) - (1-t) \BC_2(\rho_a)||_1).
    \label{prodMax1}
\end{equation}
Using 
\begin{equation}
    \DC := t \BC_1 - (1- t) \BC_2,
    \label{diffMap}
\end{equation}
one may rewrite the above expression,
\begin{equation}
    q^* = \frac{1}{2} (1 + \max_{\rho_a}|| \DC(\rho_a) ||_1).
    \label{prodMax}
\end{equation}
One can show that~(see discussion below Def.~3.37 in~\cite{Watrous18}) 
\begin{equation}
    \max_{\rho_a}|| \DC(\rho_a) ||_1 = ||\DC||_1,
    \label{prodMax2}
\end{equation}
where the 1-norm $||\DC||_1$ of the map $\DC$ is the maximum
1-norm of the operator $||\DC(X)||_1$ where $||X||_1 \leq 1$.

In the protocol above, instead of preparing a state $\rho_a$, Alice can prepare
a~(possibly entangled) state $\rho_{ar}$ on $\HC_a \ot \HC_r$, tensor product
of $\HC_a$ with $\HC_r$~(see discussion containing~\eqref{tenProd1}
and~\eqref{tenProd5}), a space of some auxiliary system with dimension $d_r$ of
Alice's choosing.  Alice can then send the $a$ sub-system of $\rho_{ar}$ to
Bob. Bob applies the map $\BC_i$ to this sub-system and returns the sub-system
to Alice.  The final state with Alice is given by $\BC_i \ot \IC_r
(\rho_{ar})$, where $\IC_r$ is the identity channel on $\HC_r$. By varying
$\rho_{ar}$ and $d_r$, the maximum probability with which Alice can correctly
predict $i$ by measuring $\BC_i \ot \IC_r (\rho_{ar})$ is
\begin{equation}
    s^* = \frac{1}{2} (1 + \sup_{\rho_{ar}, d_r}||\DC \ot \IC_r (\rho_{ar}) ||_1).
    \label{entMax}
\end{equation}
One can show that~~(see discussion below Def.~3.43 in~\cite{Watrous18})
\begin{equation}
     \sup_{\rho_{ar}, d_r}|| \DC \ot \IC_r(\rho_{ar}) ||_1 = ||\DC ||_{\diamond},
    \label{entMax2}
\end{equation}
where the diamond norm~(also called the completely bounded trace norm) $|| \DC
||_{\diamond}:= || \DC \ot \IC_{a}||_1$. The equality above demonstrates that the
dimension of the auxiliary system $\HC_r$ chosen by Alice does not need to be
larger than $d_a$, the dimension of system $a$ being sent from Alice to Bob.
In addition, since $1/2 \leq s^* \leq 1$, $|| \DC ||_{\diamond}$ for
any $\DC$ of the form~\eqref{diffMap} is at most 1.

The probabilities $q$ and $s$, defined in~\eqref{prodMax} and~\eqref{entMax},
respectively, can be obtained from the $1$-norm and the diamond norm of the map
$\DC$ defined in~\eqref{diffMap}.  As mentioned below~\eqref{entMax2}
and~\eqref{prodMax2}, for any map $\DC : \hat \HC_a \to \hat \HC_b$, both
the $1$-norm $||\DC||_1$ and the diamond norm $|| \DC ||_{\diamond}$ are
defined as the maximum value of some convex function.  Computing the maximum of
a convex function is a non-trivial problem. However, in the special case of the
diamond norm, such a maximization can be reframed as a semi-definite program.
Consider a linear superoperator $\DC: \hat \HC_a \to \hat \HC_b$ with
Choi-Jamio\l{}kowski representation $\JC_{ba}(\DC)$~(for
definition, see~\eqref{ChoiJ}). The diamond norm of $\DC$ is the optimal value
of the 
    \begin{equation}
        \text{Primal SDP:}
        \label{primalDiamond}
    \end{equation}
    \begin{align*}
        \begin{aligned}
            \text{maximize}\; & \frac{1}{2} \big(\Tr(\JC_{ba}(\DC)X) + \Tr(\JC_{ba}(\DC)X)^* \big) &\\
            \text{subject to} \;&
            \begin{pmatrix}
                I_b \ot \rho_a & X \\ 
                X^{\dag} & I_b \ot \sg_a
            \end{pmatrix}
            \succeq 0, & \\ 
            &\Tr(\rho_a) = 1,& \\
            &\Tr(\sg_a) = 1, & \\
            \text{and} \; & X \in \LC(\HC_a),  &
        \end{aligned}
    \end{align*}
and
    \begin{equation}
        \text{Dual SDP:}
        \label{dualDiamond}
    \end{equation}
    \begin{align*}
    \begin{aligned}
        \text{minimize}\; & \frac{1}{2}\big( ||\Tr_{b}(N_{ab})||_{\infty} +  ||\Tr_{b}(M_{ab})||_{\infty} \big) & \\
        \text{subject to} \; & 
        \begin{pmatrix}
            N_{ba} & -\JC_{ba}(\DC) \\
            -\JC_{ba}(\DC)^{\dag} & M_{ba}
        \end{pmatrix} 
        \succeq 0, & 
    \end{aligned}
    \end{align*}

where $||A||_{\infty}$ denotes the infinity norm, \textcolor{black}{also called
the spectral norm} of the operator $A \in \hat \HC$~\cite{Watrous09}.

This norm is \textcolor{black}{dual to the nuclear norm. It is} the minimum real $\mu$ for which $A \preceq \mu I$. Using this
characterization of the infinity norm, the dual semi-definite program above can
be re-written as
\begin{align}
    \begin{aligned}
        \text{minimize}\; &  \frac{1}{2}( \mu + \nu) & \\
        \text{subject to}\; &
    \begin{pmatrix}
        N_{ba} & -\JC_{ba}(\DC) \\ \nonumber
        -\JC_{ba}(\DC)^{\dag} & M_{ba}
    \end{pmatrix}
        \succeq 0, & \\
        & \Tr_{b}(N_{ba}) \preceq  \mu I_a, &\\
        & \Tr_{b}(M_{ba}) \preceq \nu I_a, & \\ 
        \text{and} \; & \mu, \nu \in \Rbb &.
    \end{aligned}
\end{align}
 
\textcolor{black}{These SDPs above can be derived using a connection between the
fidelity function and the diamond norm. The connection, together with a
simplification of the SDP~\eqref{SDPFid} for the fidelity function, can be used
to arrive at the above SDP~(for details see~\cite{Watrous12}).}

Let us consider some simple examples where the diamond norm can be computed
algebraically in closed form.  Our first example is the diamond norm of any
quantum channel $\BC : \hat \HC_a \to \hat \HC_b$. Using the definition of
the diamond norm $||\BC ||_{\diamond}$, one can easily show that any quantum
channel has diamond norm one. 
Our second example considers the protocol between Alice and Bob discussed at
the beginning of this section.  In that protocol, suppose $d_a = d_b = 2$; in
other words, Alice and Bob exchange qubit states. In addition, consider a
concrete case where $\BC_1$ is simply the identity channel $\IC$ and $\BC_2$ is
the qubit depolarizing channel $\Dl$ in~\eqref{depol}, and each is
equally probable, i.e., $t = 1/2$; thus $\DC$ in~\eqref{diffMap} takes
the form
\begin{equation}
    \DC(\rho) = \frac{1}{2}\big(\BC_1(\rho) - \BC_2(\rho) \big) = \frac{(1-\lm)}{2}(\rho - \Tr(\rho)\frac{I}{2}).
    \label{DForm}
\end{equation}
Using the form of $\DC$ in~\eqref{DForm}, we find~\textcolor{black}{(see App.~\ref{AppNorm} for details)}
\begin{equation}
    ||\DC||_1 = (1-\lm)/2, \quad \text{and} \quad ||\DC||_{\diamond} = 3(1-\lm)/4.
    \label{depolNorm}
\end{equation}
The norms above, along with equations~\eqref{prodMax} and~\eqref{entMax},
give
\begin{equation}
    q^* = (3-\lm)/4 \quad \text{and} \quad s^* = (7-3\lm)/8 > q^*.
    \label{qsVals}
\end{equation}
Since $s^*$ is larger than $q^*$, we find that Alice can increase the
probability of correctly distinguishing $\Delta$ from $\IC$ using a qubit
auxiliary system.  This increase comes from the possibility of Alice sending to
Bob one half of a joint system, in state $\rho_{aa}$. Later, system $b$
returned by Bob to Alice results in a joint state $\rho_{ba}$ which Alice may
measure jointly.

Consider an extreme case where $\lm = -1/3$~\cite{Watrous08}.  Suppose Alice
prepares a pure entangled state, $\rho_{aa} = \dya{\psi^+}$, and
\begin{equation}
    \ket{\psi^+} = \frac{1}{\sqrt{2}}(\ket{00} + \ket{11}).
\end{equation}
One-half of this entangled state is sent to Bob. If Bob applies the identity
channel $\IC$, the state $\rho_{ba}$ remains unchanged. If Bob applies
$\Delta$, the state $\rho_{ba}$ becomes orthogonal to $\dya{\psi^+}$.  This
orthogonal state can be distinguished perfectly by doing a joint measurement on
$ba$~(see discussion accompanying~\eqref{eq:piarOr}).
As a result, we get $s^* = 1$. Without using such entangled inputs, the
maximum probability of distinguishing $\IC$ from $\Delta$ is $q^* = 5/6$. 
We test these findings numerically. In particular, we compute the diamond norm
of $\DC$ in~\eqref{DForm} by solving the primal and dual
SDPs,~\eqref{primalDiamond} and~\eqref{dualDiamond}. The optimum value of both
SDPs is in good agreement, and their difference is typically $O(10^{-10})$.  Using
this value, we obtain the numerical estimate $s^*_n$~\eqref{entMax} of
$s^*$~\eqref{qsVals}; their absolute difference is typically small
$O(10^{-11})$ too. Comparing $s^*_n$ with $q^*$ reveals $s^*_n$ is larger
and thus entanglement helps discriminate quantum channels $\IC$ and $\Dl$.

{\bf Remark.} The above example can be generalized to a $d$-dimensional quantum
system called a {\em qudit}. In this higher-dimensional case, $\BC_1$ is the
qudit identity channel $\IC_d$ and $\BC_2$ is the qudit depolarizing channel,
\begin{equation}
    \Delta_d(\rho) = \lm \rho + (1-\lm) \Tr(\rho) \frac{I}{d}.
    \label{eq:quditDepol}
\end{equation}
In~\eqref{diffMap} setting $t = 1/2,\BC_1 = \IC_d$ and $\BC_2 = \Delta_d(\rho)$
results in a superoperator 
\begin{equation}
    \DC (\rho) = \frac{1-\lm}{2} (\rho - \Tr(\rho) \frac{I}{d}),
    \label{eq:del}
\end{equation}
whose $1$ norm and diamond norm are~(see App.~\ref{AppNorm} for details)
\begin{equation}
    ||\DC||_1 = (1-\lm)\frac{d-1}{d}, \quad \text{and} \quad ||\DC||_{\diamond} = (1-\lm)\frac{d^2-1}{d^2},
    \label{DcVals}
\end{equation}
respectively.
Using the above equations, along with \eqref{prodMax} and \eqref{entMax}, one finds
that $s^* > q^*$ and the difference $s^* - q^*$ scales as $O(1/d)$.

In our third channel discrimination example~(based on Ex.~3.36
in~\cite{Watrous18}), we again consider the protocol between Alice and Bob
discussed at the beginning of this section.  In this protocol Alice and Bob
exchange qudits.  With equal probability, i.e., $t = 1/2$, Bob chooses one of
two Werner-Holevo channels
\begin{equation}
    \BC_1(\rho) = \frac{1}{d+1}\big(\Tr(\rho) I + \rho^T \big) 
    \quad \text{or} \quad
    \BC_2(\rho) = \frac{1}{d-1}\big(\Tr(\rho) I - \rho^T \big),
\end{equation}
where the transpose is done in the standard basis. Channels $\BC_1$ and $\BC_2$
have Choi-Jamio\l{}kowski representations,
\begin{equation}
    \JC_{ba}(\BC_1) =  \frac{1}{d+1}( I_b \otimes I_a + S_{ba}), 
    \quad \text{and} \quad
    \JC_{ba}(\BC_2) = \frac{1}{d-1}( I_b \otimes I_a - S_{ba}), 
\end{equation}
respectively, where 
\begin{equation}
    S_{ba} \ket{i} \ot \ket{j}= \ket{j} \ot \ket{i}
    \label{eq:swap}
\end{equation}
is the swap operator acting on $\HC_b \ot \HC_a$~(spaces $\HC_a$ and $\HC_b$
have equal dimension $d$).  The operators $(I_b \otimes I_a + S_{ba})/2$ and
$(I_b \otimes I_a - S_{ba})/2$ equal the projector onto the symmetric and
anti-symmetric sub-spaces of $\HC_b \ot \HC_a$, respectively. For these
Werner-Holevo channels, $\DC$ in~\eqref{diffMap} takes the form
\begin{equation}
    \DC(\rho) = \frac{d}{d^2 - 1} \big( \rho^T - \Tr(\rho) \frac{I}{d}\big).
    \label{DiffWH}
\end{equation}
The operator norm $||\DC||_1 =  2/(d+1)$; using~\eqref{prodMax1}, we get 
\begin{equation}
    q^* = \frac{1}{2} + \frac{1}{d+1}.
    \label{pStarWernerH} 
\end{equation}
As stated below~\eqref{entMax}, $||\DC||_{\diamond}$ is at most 1 and equals
the maximum value of $|| \DC \ot \IC (X)||_1$ where $||X||_1 \leq 1$.  Let $X$
be a projector onto the maximally entangled states on two qudits,
\begin{equation}
    \ket{\phi_d} := \frac{1}{\sqrt{d}} \sum_i \ket{i}_a \ot \ket{i}_a,
    \label{maxEntd}
\end{equation}
then
\begin{equation}
    || \DC \ot \IC (X)||_1 = || \frac{1}{d} \big( \JC(\BC_1)_{ba} - \JC(\BC_2)_{ba} \big)|| = 1.
\end{equation}
As a result, $||\DC||_{\diamond} = 1$. From~\eqref{entMax} and~\eqref{entMax2}
it follows that 
\begin{equation}
    s^* = 1.
\end{equation}
A numerical estimate $s^*_n$ of $s^*$ is found by directly using the definition
of $s^*$~\eqref{entMax} and by solving the primal and dual
SDPs,~\eqref{primalDiamond} and~\eqref{dualDiamond} with $\DC$
in~\eqref{DiffWH} and $d = 3$. This numerical estimate is in reasonably
good agreement with the value of $s^*$ stated above.

Notice that for any $d$, there is a gap between $s^*$ above and $q^*$
in~\eqref{pStarWernerH} which is at least $1/2$ and scales as $O(1/d)$. The
strategy that allows us to obtain $s^* = 1$ makes use of entanglement. In
particular the maximally entangled state $\dya{\phi_d}$~\eqref{maxEntd} can be
sent by Alice to Bob. If Bob applies $\BC_1$ then Bob's joint state with the
auxilliary system is $\JC_{ba}(\BC_1)/\Tr(\JC_{ba}(\BC_1))$, otherwise Bob
applies $\BC_2$, creating the joint state
$\JC_{ba}(\BC_2)/\Tr(\JC_{ba}(\BC_2))$. Notice these two joint states are
orthogonal to each other, and thus can be distinguished perfectly by doing a
joint measurement on $\HC_{ba}$~(see discussion
accompanying~\eqref{eq:piarOr}). Notice that measuring the joint system is
necessary, if Alice throws away the reference system $a$ and keeps only $b$;
then the states received by Alice from Bob would be identical, regardless of
the channels $\BC_1$ and $\BC_2$ applied by Bob.

In the special cases discussed above, we provided algebraic expressions for the
diamond norm. When such algebraic expressions are not available, one needs to
use numerical techniques to a solve semi-definite program and obtain the
diamond norm. A short tutorial on solving these semi-definite programs using
Python packages PICOS and solvers in CVXOPT/MOSEK is available along with this
chapter~(see Notebook~3 in~\cite{NoName21a}).

The SDP formulation of the diamond norm has several virtues. It shows that a
variety of bounds on the ability of quantum channels to send information can be
computed efficiently~\cite{HolevoWerner01,WangDuan16}. Not only bounds, one
can use the SDP formulation to obtain computable measures of
entanglement~\cite{WangWilde18}, a non-trivial problem in entanglement theory.
The SDP formulation has also aided the use of diamond norm in quantum error
correction~\cite{IyerPoulin17, MagesanGambettaEA12} and compressed
sensing~\cite{KlieschKuengEA16}.

\section{Problem 4: Quantum Entanglement and Separability}
\label{P4EntSep}

Entanglement is a patently non-classical aspect of quantum states. Any bipartite
quantum state on $\HC_a \ot \HC_b$ is said to be entangled if the state's
density operator $\rho_{ab}$ cannot be expressed as a convex combination,
\begin{equation}
    \rho_{ab} = \sum_i p_i \dya{\psi_i} \ot \dya{\phi_i},
    \label{sepState}
\end{equation}
of pure product states $\ket{\psi_i} \ot \ket{\phi_i}$ on the tensor
product~(see discussion containing~\eqref{tenProd5} for the definition) of $\HC_a$
and $\HC_b$; here $p_i \geq 0$ and $\sum_i p_i = 1$. A quantum state with a
density operator of the form in~\eqref{sepState} is called a separable state.
In what follows, we discuss criteria for the separability of quantum states.
When $\rho_{ab}$ is a pure state, i.e., $\rho_{ab}^2 = \rho_{ab}$, then
criteria for separability of $\rho_{ab}$ is relatively easy~(see discussion at
the end of Sec.~\ref{SStatesDirac}). To discuss a criterion of separability that
applies more generally, we need a notion called the {\em extension} of a density
operator.

Consider a density operator $\rho_{ab}$ on $\HC_{ab}$. For
any integer $n \geq 1$, let $\HC_{bk}$ be a space of a dimension equal
to that of $\HC_{b}$ and 
\begin{equation}
    \HC_B := \bigotimes_{j=1}^k \HC_{bj},
    \label{BDef}
\end{equation}
where $\bigotimes$ is our notation for tensor product~(defined in the discussion
containing~\eqref{tenProd1} and~\eqref{tenProd2}) of several spaces $\HC_{b1},
\HC_{b2}, \cdots ,\HC_{bk}$.
For $k=1$, $\HC_B = \HC_{b1}$ and a density operator $\rho_{aB}$ on $\HC_{aB}$
is an extension of $\rho_{ab}$ if $\rho_{aB} = \rho_{ab}$. For $k>1$,
$\rho_{aB}$ is called an extension of $\rho_{ab}$ if the partial trace
$\rho_{ab1}$ of $\rho_{aB}$ over $\HC_{b2} \ot \HC_{b3} \ot \cdots \ot \HC_{bk}$
is $\rho_{ab}$; in other words,
\begin{equation}
    \rho_{ab} = \rho_{ab1} = \Tr_{b2,b3,\dots ,bn}(\rho_{aB}).
    \label{ext}
\end{equation}
An extension $\rho_{aB}$ of $\rho_{ab}$ is called symmetric if swapping any
space $\HC_{bi}$ with $\HC_{bj}$ in $\HC_B$, where $1 \leq i < j \leq n$, has
no effect on $\rho_{aB}$; that is,
\begin{align}
    \rho_{aB} = \Pi_{bibj} \rho_{aB} \Pi_{bibj},
    \label{symm}
\end{align}
where the swap operator $\Pi_{bibj}$ on $\HC_a \ot \HC_B$ simply extends the
usual swap operator $S_{bibj}$~\eqref{eq:swap} on $\HC_{bi} \ot
\HC_{bj}$ to $\HC_a \ot \HC_B$. This extension applies $S_{bibj}$ to $\HC_{bi}
\ot \HC_{bj}$ and the identity to all other spaces.

A quantum state $\rho_{ab}$ is separable if and only if it has a symmetric
extension for all $k \geq 1$~\cite{FannesLewisEA88, RaggioWerner89,Werner89}.
Further, one can show that if $\rho_{ab}$ has a symmetric extension for some
$k$, then it has a symmetric extension for all $k' > k$.  These powerful
symmetric extension results provide a straightforward route to check if a given
density operator $\rho_{ab}$ is separable or entangled.  This route is to pick
an integer $k \geq 1$ and frame a constraint satisfaction problem which simply
checks if the linear constraints~\eqref{ext} and~\eqref{symm} can be satisfied
by a unit-trace positive semi-definite operator $\rho_{aB}$. For each $k$, this
problem can also be framed as a SDP.  If this SDP is
infeasible, then $\rho_{ab}$ does not have a symmetric extension for that $k$
and hence $\rho_{ab}$ must be entangled.  On the other hand if the SDP is
feasible, then $\rho_{ab}$ may still be entangled.

Another criterion for checking if a state $\rho_{ab}$ is separable is the
positive under partial transpose~(PPT) criterion~\cite{Peres96}. Notice
$\TC_b$, a transpose with respect to $\HC_b$ of the separable state of
$\rho_{ab}$ in~\eqref{sepState}, results in
\begin{equation}
    \TC_{b} (\rho_{ab}) = \sum_i p_i \dya{\psi_i} \ot \dya{\phi_i}^T,
    \label{sepStateTr}
\end{equation}
where the superscript $T$ represents transpose in the standard basis of
$\HC_b$. For any separable state $\rho_{ab}$, the operator $\TC_{b}
(\rho_{ab})$ above is positive semi-definite; thus, one says any separable
state $\rho_{ab}$ is PPT. On the other hand, if for some state $\sg_{ab}$ the
partial transpose $\TC_{b} (\sg_{ab})$ is not PPT, then $\sg_{ab}$ cannot be
separable; that is, $\sg_{ab}$ is entangled if it is not PPT. For $d_a =2$ and
$d_b = 2,3$, this PPT criterion is both necessary and
sufficient~\cite{HorodeckiHorodeckiEA96}. But generally, this PPT criterion is
necessary but not sufficient.  However, this PPT criterion can be combined with
the previously discussed necessary and sufficient condition for a separable
state to have a symmetric extension for all $k$.  This combination results in
the {\em PPT symmetric extension} criterion for
separability~\cite{DohertyParriloEA04}.  In this criterion, for any $k \geq 1$,
the symmetric extension $\rho_{aB}$ of $\rho_{ab}$ must also be PPT, where the
partial transpose is taken with respect to each of the $k$ spaces $\HC_{b1}$,
$\HC_{b1 b2}$, $\dots$, $\HC_{b1 b2 \dots bk}$; that is,
\begin{equation}
    \TC_{b1 b2 \dots bj} (\rho_{aB}) \succeq 0,
    \label{sepStateTra}
\end{equation}
for all $1 \leq j \leq k$. 

The PPT symmetric extension criterion also provides a route to check if a
density operator $\rho_{ab}$ is separable or entangled. In this route, one
picks an integer $k$ and first formulates an SDP for finding a symmetric
extension $\rho_{aB}$ and then adds linear PPT constraints~\eqref{sepStateTra}
to this SDP. These additional PPT constraints can turn an SDP that was feasible
to one that isn't. Such an infeasible SDP indicates the absence of a PPT
symmetric extension, and thus the presence of entanglement. As a result, the
PPT symmetric extension-based route to checking entanglement can be strictly
better than the usual symmetric extension route. 

The SDP formulations discussed so far were about checking feasibility. From a
numerical standpoint, it is convenient to reformulate these feasibility
problems as SDP optimization problems. For any $k \geq 1$, the SDP optimization
problems arising from the PPT symmetric extension criterion can be written as
\begin{align}
    \label{SDPRepPPTEx}
    \begin{aligned}
    \text{minimize} \; &   \mu \\ 
        \text{subject to} \; & \rho_{ab} = \Tr_{b2,b3,\dots bk}(\rho_{aB}), &  \\ 
        & \rho_{aB} = \Pi_{bibj} \rho_{aB} \Pi_{bibj},& &1 \leq i < j \leq k,  \\
        & \Tr(\rho_{aB}) = 1,&  \\
        & \rho_{aB} + \mu I_{aB} \succeq 0, & \\
        \text{and} \; & \TC_{b1 b2 \dots bj} (\rho_{aB}) + \mu I_{aB} \succeq 0, & &1 \leq j \leq k .&  
    \end{aligned}
\end{align}
Dropping the last constraint in the SDP above results in an SDP arising from
the symmetric extension criterion. For some fixed $k$, if $\rho_{ab}$ does not
have a PPT symmetric extension then the optimum value of the above SDP,
$\mu^*$, is strictly positive. This strict positivity implies that $\rho_{ab}$
is entangled.  On the other hand, if $\mu^*$ is zero or less than zero, then
$\rho_{ab}$ has a PPT symmetric extension and this $\rho_{ab}$ may or may not
be entangled.

The simplest example of the SDP in~\eqref{SDPRepPPTEx} occurs for $k=1$. In
this case $\HC_B = \HC_{b1} = \HC_b$, and thus for any state $\rho_{ab}$, 
the SDP above can be reduced to the form
\begin{align}
    \label{SDPRepPPTEx1}
    \begin{aligned}
    \text{minimize} \; &   \mu \\ 
        \text{subject to} \; & \TC_{b} (\rho_{ab}) + \mu I_{ab} \succeq 0.
    \end{aligned}
\end{align}
Here, $\mu^*$ is simply obtained by putting a negative sign in front of the
smallest eigenvalue of $\TC_{b} (\rho_{ab})$, the partial transpose of
$\rho_{ab}$ with respect to $\HC_b$. Suppose $d_a = d_b = 2$, and $\rho_{ab} =
\dya{\chi}$, where $\ket{\chi}$ is defined in~\eqref{maxEnt}; then one may
write
\begin{equation}
    \rho_{ab} = \frac{1}{2} (\dya{0}_a \ot \dya{0}_b
    + \dyad{1}{0}_a \ot \dyad{1}{0}_b
    + \dyad{0}{1}_a \ot \dyad{0}{1}_b
    + \dya{1}_a \ot \dya{1}_b,
    ),
\end{equation}
or express $\rho_{ab}$ above as a $4 \times 4$ matrix in the $\ket{i}_a \ot
\ket{j}_b$ basis as
\begin{equation}
    \rho_{ab} = \frac{1}{2}
    \begin{pmatrix}
    1 & 0 & 0 & 1 \\
    0 & 0 & 0 & 0 \\
    0 & 0 & 0 & 0 \\
    1 & 0 & 0 & 1 
    \end{pmatrix}.
\end{equation}
The partial transpose of $\rho_{ab}$ with respect to $\HC_b$,
\begin{equation}
    \TC_b(\rho_{ab}) = \frac{1}{2} (\dya{0}_a \ot \dya{0}_b
    + \dyad{1}{0}_a \ot \dyad{0}{1}_b
    + \dyad{0}{1}_a \ot \dyad{1}{0}_b
    + \dya{1}_a \ot \dya{1}_b
    ),
\end{equation}
can also be written in a matrix form,
\begin{equation}
    \TC_b(\rho_{ab}) = \frac{1}{2}
    \begin{pmatrix}
    1 & 0 & 0 & 0 \\
    0 & 0 & 1 & 0 \\
    0 & 1 & 0 & 0 \\
    0 & 0 & 0 & 1 
    \end{pmatrix},
\end{equation}
again using the $\ket{i}_a \ot \ket{j}_b$ basis.  The smallest eigenvalue of
the above matrix is $-1/2$. As a result, the optimum value of the
SDP~\eqref{SDPRepPPTEx1}, $\mu^*$, is $1/2$. A strictly positive $\mu^*$ value indicates that
$\rho_{ab} = \dya{\chi}$ is an entangled state.  This entanglement was already
discussed below~\eqref{maxEnt}, the SDP approach above merely confirms this
fact. Next, let $d_a = d_b = 3$ and $\rho_{ab}$ be a state described
in~\cite{HorodeckiHorodeckiEA01},
\begin{equation}
    \rho_{ab} = \frac{2}{7} \dya{\psi^+} + \frac{\al}{7} \sg_+ + \frac{5 - \al}{7} S_{ab} \sg_+ S_{ab},
    \label{alQ}
\end{equation}
where $0 \leq \al \leq 5$, $\ket{\psi^+} = \frac{1}{\sqrt{3}} (\ket{00} +
\ket{11} + \ket{22})$, $\sg_+ = \frac{1}{3}(\dya{01} + \dya{12} + \dya{20})$,
and $S_{ab}$ is the swap operator in~\eqref{eq:swap}. Replacing $\al$ with $5 -
\al$ in~\eqref{alQ} is equivalent to swapping the $\HC_a$ and $\HC_b$ spaces.
Such a swap has no effect on the solution to the SDP~\eqref{SDPRepPPTEx1} and
does not change whether $\rho_{ab}$ is entangled or separable.  Thus, we
restrict ourselves to $0 \leq \al \leq 5/2$. Using a procedure similar to the
one described above, one can compute $\TC_b(\rho_{ab})$ and its smallest
eigenvalue, $\al^* = (5 - \sqrt{4 \al^2 - 20 \al + 41})/42$. We know $\mu ^* =
- \al^*$. It is easy to check that $\mu^*$ is strictly positive for $0 \leq \al
< 1$, and negative for $1 \leq \al \leq 5/2$. Thus we conclude, that for $\al
<1$, the qudit state $\rho_{ab}$ in~\eqref{alQ} is entangled. For other values
of $1 \leq \al \leq 5/2$, $\mu^*$ is negative and one cannot conclude if
$\rho_{ab}$ is entangled or separable. A path forward to check entanglement for
these other values of $\al$ is to solve a larger SDP by setting $k=2$
in~\eqref{SDPRepPPTEx}. 

For $k=2$, $\HC_B = \HC_{b1} \ot \HC_{b2}$, $\rho_{ab1} = \rho_{ab}$, and the
SDP in~\eqref{SDPRepPPTEx} can be reduced to the form
\begin{align}
    \label{SDPRepPPTEx2}
    \begin{aligned}
    \text{minimize} \; &   \mu \\ 
        \text{subject to} \; & \rho_{ab1} = \Tr_{b2}(\rho_{aB}), &  \\ 
        & \rho_{aB} = \Pi_{b1b2} \rho_{aB} \Pi_{b1b2},&  \\
        & \Tr(\rho_{aB}) = 1,&  \\
        & \rho_{aB} + \mu I_{aB} \succeq 0, & \\
        & \TC_{b1} (\rho_{aB}) + \mu I_{aB} \succeq 0, &\\
        \text{and} \; & \TC_{b1 b2} (\rho_{aB}) + \mu I_{aB} \succeq 0. & 
    \end{aligned}
\end{align}
To solve this SDP, we use open-source numerical packages. A short
tutorial using Python packages PICOS and CVXOPT is available along with this
chapter~(see Notebook~4 in~\cite{NoName21a}).
Using these packages, for the above SDP we find the optimal value $\mu^*$. This
value is strictly positive for $1 \leq \al < 2$ and thus $\rho_{ab}$
in~\eqref{alQ} is entangled for $1 \leq \al < 2$.  This entanglement was not found by the SDP
in~\eqref{SDPRepPPTEx1} and it demonstrates that the SDP for $k=2$ is strictly
better at detecting entanglement than the SDP for $k=1$.
The finding that $\rho_{ab}$ in~\eqref{alQ} is entangled for $0 \leq \al <2$ is
consistent with~\cite{HorodeckiHorodeckiEA01}, where the
entanglement of $\rho_{ab}$ was first discussed. That discussion also considers
the parameter range $2 \leq \al \leq 5/2$, where $\rho_{ab}$ in~\eqref{alQ} is
shown to be separable.

The methods for checking separability discussed here are often called an SDP
hierarchy. The hierarchy discussed here is based on the work
of~\cite{DohertyParriloEA04}.  This is not the only hierarchy; a number of
other hierarchies have been studied~\cite{NavascuesPironioEA08, BertaFawziEA16,
HarrowNatarajanEA17, HarrowNatarajanEA19}. 
While we have only provided a brief introduction, the theory of quantum
entanglement is an active area of study with a variety of open problems~(see
references in and citation to~\cite{HorodeckiHorodeckiEA09a}).

\section{Problem 5: Quantum Channel Capacity}
\label{P5Cap}

Information is processed via physical media. However, because physical media
introduce noise, it is natural to ask the amount of noiseless information that
can be sent across some noisy medium. To answer such questions, one constructs
an abstract model for the noisy medium. This model is called a noisy
communication channel.  A classical channel sends distinguishable input symbols
to distinguishable output symbols.  Suppose the input symbols come from a
discrete set, an alphabet $\XC$, and the output symbols come from a possibly
different alphabet $\YC$. A discrete memoryless channel $N$ takes an input $x
\in \XC$ to an output $y \in \YC$ with probability $p(y|x)$. The input and
output may be considered random variables $X$ and $Y$, respectively, and the
channel simply takes $X$ to $Y$. This channel $N$ is called memoryless because
any output $Y=y$ only depends on the current channel input $X=x$ and not on a
prior input. 

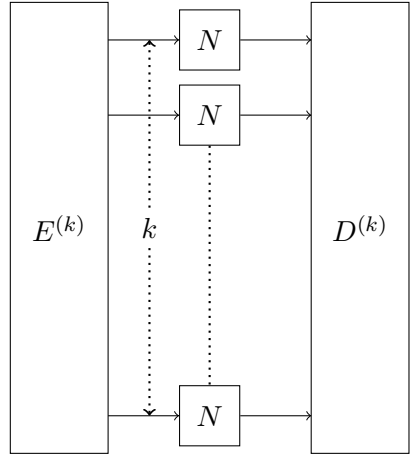
\begin{figure}
    \centering
    
    \tikzstyle{int}=[circle,draw, minimum size=3.25em]
    \tikzstyle{chan1}=[rectangle,draw,  minimum width=1.3cm, minimum height=6cm]
    \tikzstyle{chan2}=[rectangle,draw,  minimum width=.8cm, minimum height=.8cm]
    \tikzstyle{txBox} = [draw=none, fill=none, minimum height=1em, minimum width=1em]

    \begin{tikzpicture}[scale=1, every node/.style={scale=1.00}]
        \node [chan1] at (2,0) (b) {$E^{(k)}$};

        \node [txBox] at (3.2,0) (c) {$k$};
        \node [chan2] at (4,2.5) (c1) {$N$};
        \node [chan2] at (4,1.5) (c3) {$N$};
        \node [chan2] at (4,-2.5) (c2) {$N$};
        
        \node [chan1] at (6,0) (d) {$D^{(k)}$};

        \draw[->, dotted, thick] (c) -- (3.2,2.5) ;
        \draw[->, dotted, thick] (c) -- (3.2,-2.5) ;
        
        \draw[->] (2.65,2.5) -- (c1) ;
        \draw[->] (2.65,1.5) -- (c3) ;
        \draw[->] (2.65,-2.5) -- (c2) ;
        \draw[->] (c1) -- (5.35,2.5) ;
        \draw[->] (c3) -- (5.35,1.5) ;
        \draw[->] (c2) -- (5.35,-2.5) ;

        \path[dotted, thick] (c3) edge              node  {} (c2);

    \end{tikzpicture}
    \caption{Encoding, $E^{(k)}$, and decoding, $D^{(k)}$, classical information
    across $k$ uses of a classical channel $N$.}
    \label{fig:achCl}
\end{figure}

Noise introduced by a channel can be corrected by error-correcting codes that
encode and decode information across multiple channel uses.  The rate of error
correction across multiple channel uses is captured by the notion of an {\em
achievable rate}. Roughly speaking, an encoding $E^{(k)}$ and decoding
$D^{(k)}$ over $k$ joint uses of a channel $N$~(see Fig.~\ref{fig:achCl}) that sends $kR$ bits with
vanishing error as $k \mapsto \infty$ is said to have an achievable rate
$R$. The maximum possible achievable rate is called
the channel capacity $C(N)$.  Achievable rates and channel capacity are
fundamental quantities in information theory; Shannon~\cite{Shannon48} provided
a simple way to compute an achievable rate for any given channel $N$ with
conditional probability $p(y|x)$. This rate, which we call the {\em channel
mutual information}, is simply given by
\begin{equation}
    C^{(1)}(N) = \max_{p(x)} I(X:Y),
    \label{chMI}
\end{equation}
where $p(x)$ is a probability distribution over input symbols $x$, and $I(X;Y)$
is the mutual information~\eqref{mi} between the input $X$ and output $Y$. For
any fixed $N$ ---that is, fixed $p(y|x)$--- the mutual information is concave
in $p(x)$~(see Th.2.7.4 in~\cite{CoverThomas01}). As a result, $C^{(1)}(N)$ can be computed efficiently using tools
from convex optimization~\cite{Blahut72, Arimoto72}. In addition, the channel
mutual information is additive: for any two channels $N$ and $N'$ used together
the channel mutual information $C^{(1)}(N \times N')$ is simply the sum
$C^{(1)}(N) + C^{(1)}(N')$.
The channel capacity $C(N)$ can be written in terms of the
channel mutual information as a limit
\begin{equation}
    C(N) = \lim_{k \mapsto \infty} \frac{1}{k} C^{(1)}(N^{\times k}),
    \label{chCap1}
\end{equation}
where $N^{\times k}$ represents $k$ joint uses of $N$. This limit greatly
simplifies due to additivity, $C(N) = C^{(1)}(N)$, a remarkable {\em
single-letter} expression. From this expression, the capacity of any channel
$N$ to send information over infinitely many channel uses is given by the
maximum mutual information between the input and output of a single use of the
channel.

Both information and the physical medium carrying this information can be
modelled using quantum mechanics.  In this model, noise introduced by physical
medium is described by a quantum channel. A fundamental question in quantum
information theory is to understand the maximum rate at which noiseless quantum
information can be sent across a noisy quantum channel~(for instance
see~\cite{BennettShor98}).  This question is answered in a way analogous to the
one above used by Shannon to understand the capacity of a classical channel.

\begin{figure}
    \centering
    
    \tikzstyle{int}=[circle,draw, minimum size=3.25em]
    \tikzstyle{chan1}=[rectangle,draw,  minimum width=1.3cm, minimum height=6cm]
    \tikzstyle{chan2}=[rectangle,draw,  minimum width=.8cm, minimum height=.8cm]
    \tikzstyle{txBox} = [draw=none, fill=none, minimum height=1em, minimum width=1em]

    \begin{tikzpicture}[scale=1, every node/.style={scale=1.00}]
        \node [chan1] at (2,0) (b) {$\EC^{(k)}$};

        \node [txBox] at (3.2,0) (c) {$k$};
        \node [chan2] at (4,2.5) (c1) {$\BC$};
        \node [chan2] at (4,1.5) (c3) {$\BC$};
        \node [chan2] at (4,-2.5) (c2) {$\BC$};
        
        \node [chan1] at (6,0) (d) {$\DC^{(k)}$};

        \draw[->, dotted, thick] (c) -- (3.2,2.5) ;
        \draw[->, dotted, thick] (c) -- (3.2,-2.5) ;
        
        \draw[->] (2.65,2.5) -- (c1) ;
        \draw[->] (2.65,1.5) -- (c3) ;
        \draw[->] (2.65,-2.5) -- (c2) ;
        \draw[->] (c1) -- (5.35,2.5) ;
        \draw[->] (c3) -- (5.35,1.5) ;
        \draw[->] (c2) -- (5.35,-2.5) ;

        \path[dotted, thick] (c3) edge              node  {} (c2);

    \end{tikzpicture}
    \caption{Encoding, $\EC^{(k)}$, and decoding, $\DC^{(k)}$, quantum information
    across $k$ uses of a quantum channel $\BC$.}
    \label{fig:achQm}
\end{figure}

Noise introduced by a quantum channel $\BC$ can be corrected by using quantum
error-correcting codes that encode and decode information across quantum
channels. Roughly speaking, a quantum code with
encoding $\EC^{(k)}$ and decoding $\DC^{(k)}$ over $k$ joint uses of a channel
$\BC$~(see Fig.~\ref{fig:achQm}), which sends $kR$ qubits with vanishing error as $k \mapsto \infty$, is said to
have an achievable rate $R$. The maximum possible achievable rate of this type
is defined to be the quantum capacity $\QC(\BC)$.  The quantum analog of the
channel mutual information $C^{(1)}$ is the {\em channel coherent information} of a
channel $\BC:\hat \HC_a \to \hat \HC_b$~\cite{BarnumNielsenEA98},
\begin{equation}
    \QC^{(1)}(\BC) = \max_{\rho_a} I_c(\BC, \rho_a),
    \label{eq:CICal}
\end{equation}
where $I_c(\BC,\rho_a): = S\big( \BC(\rho_a) \big) - S\big( \BC^c(\rho_a)
\big)$ is the {\em entropy bias} or the {\em coherent information} of a channel
$\BC$ at $\rho$ and $\BC^c:\hat \HC_a \to \hat \HC_c$ is the complementary
channel of $\BC$ and $\HC_c$ the complementary output space, sometimes called
the environment of $\BC$. Recall any channel $\BC$ has a Kraus
decomposition~\eqref{Qchan}, written using $d_c$ Kraus operators $K_i: \HC_a
\to \HC_b$; the complementry channel $\BC^c$ can be defined using a Kraus
decomposition,
\begin{equation}
    \BC^c(A) = \sum_{j=1}^{d_b} L_j A L_j^{\dag},
    \label{eq:ChanComp}
\end{equation}
where the Kraus operators $L_j : \HC_a \to \HC_c$ have matrix elements
$[L_j]_{ki} = [K_i]_{jk}$. 

Unlike $I(X:Y)$, $I_c(\BC, \rho_a)$ is not necessarily concave in the input
$\rho_a$ for fixed $\BC$. As a result, despite the fundamental importance of
$\QC^{(1)}(\BC)$, methods for computing $\QC^{(1)}$ are limited.  While
$C^{(1)}$ is additive, its quantum analog $\QC^{(1)}$ is non-additive; that is,
for two channels $\BC$ and $\BC'$ used together, the coherent information of the
joint channel satisfies an inequality
\begin{equation}
    \QC^{(1)}(\BC \ot \BC') \geq \QC^{(1)}(\BC) + \QC^{(1)}(\BC'),
    \label{qmNon}
\end{equation}
which can be strict~\cite{DiVincenzoShorEA98, SmithSmolin07, SmithYard08,
LeditzkyLeungEA18, BauschLeditzky20, Siddhu20, BauschLeditzky21, Siddhu21},
where $\BC \ot \BC'$ is the tensor product of two channels~(see the discussion
containing~\eqref{tenProd6} for definition).

The quantum capacity $\QC(\BC)$ can be written in terms of the
channel coherent information as a limit~\cite{Lloyd97, Shor02a, Devetak05}
\begin{equation}
    \QC(\BC) = \lim_{k \mapsto \infty} \frac{1}{k} \QC^{(1)}(\BC^{\otimes k}),
    \label{qmCap1}
\end{equation}
where $\BC^{\otimes k}$ denotes $k$ tensor products of $\BC$ with itself.
This expression for the quantum capacity requires computing a limit over
multiple uses of the same channel. In general, this {\em multi-letter}
expression for $\QC$ is intractable to compute because of non-additivity. As a
result, $\QC^{(1)}$ is always a lower bound on $\QC$, but it need not equal
$\QC$. For a special class of channels called {\em degradable channels}
$\QC^{(1)} = \QC$~\cite{DevetakShor05}.  A channel $\BC$ is said to be
degradable if there is another channel $\CC$ such that $\CC \circ \BC = \BC^c$,
and $\BC^c$, the complement of $\BC$, is called anti-degradable. For any two
(anti)~degradable channels $\BC$ and $\BC'$, the inequality in~\eqref{qmNon} is
an equality~\cite{LeditzkyDattaEA18}; that is, additivity holds.  Such
additivity simplifies the multi-letter expression for $\QC$ in~\eqref{qmCap1}
to a single-letter formula: 
\begin{equation}
    \QC(\BC) = \QC^{(1)}(\BC),
    \label{addDeg}
\end{equation}
where $\BC$ is a degradable or anti-degradable channel. For a degradable
channel, $\QC^{(1)}(\BC)$ is relatively easy to compute: the entropy bias
$\Dl(\BC,\rho_a)$ for a degradable channel $\BC$ is a concave function of
$\rho$~\cite{YardHaydenEA08} and can be maximized using tools from convex
optimization~\cite{FawziFawzi18, RamakrishnanItenEA21}.

Given the role of degradable channels in simplifying the discussion of quantum
capacities, it is natural to ask if approximate notions of degradable channels
can approximately simplify the discussion of quantum capacities. One such
approximate notion is that of an $\ep$-degradable channel~\cite{SutterScholzEA17}. A channel $\BC$
is $\ep$-degradable if there is another channel $\CC$ such that
\begin{equation}
    ||\CC \circ \BC - \BC^c||_{\diamond} = \ep.
    \label{epsDeg}
\end{equation}
If $\ep$ is zero, then $\BC$ is degradable and~\eqref{addDeg} holds. If
$\ep$ is not zero, then~\eqref{addDeg} gets modified to
\begin{equation}
    \QC^{(1)}(\BC) \leq \QC(\BC) \leq 
    \QC^{(1)}(\BC) + \ep \log(d_c - 1)/2 + h(\ep/2) +\ep \log(d_c) + (1 + \frac{\ep}{2}) h\big( \ep/(2 + \ep) \big),
    \label{approxDeg}
\end{equation}
where $d_c$ is the output dimension of the complementary channel $\BC^c$.
The smallest $\ep$ for which~\eqref{epsDeg} holds, $\ep_{\BC}$, can be found using
a semi-definite program~(SDP),
\begin{align}
    \label{SDPRepEpsD}
    \begin{aligned}
    \text{minimize} \; & 2 \mu &\\ 
    \text{subject to} \; & \Tr_c (Z_{ca}) \preceq \mu I_a, & \\
        & \Tr_c\big( \JC_{cb}(\CC) \big) = I_b, & \\
        & Z_{ca} \succeq \JC_{ca}(\BC^c) - \JC_{ca}(\CC \circ \BC), & \\
        & Z_{ca} \succeq 0, & \\
        \text{and} \;& \JC_{cb}(\CC) \succeq 0. &
    \end{aligned}
\end{align}
In this SDP, $\mu$ is a real variable while $Z_{ca}$ and $\JC_{cb}(\CC)$ are
positive semi-definite variables. These variables satisfy linear constraints.
In particular the constraint $Z_{ca} \succeq \JC_{ca}(\BC^c) - \JC_{ca}(\CC
\circ \BC)$ is linear in $\JC_{cb}(\CC)$ because $\JC_{ca}(\CC \circ \BC)$ is
linear in $\JC_{cb}(\CC)$ and can be written using $\JC_{ca}(\BC)$~(see
App.~\ref{AppLin}). The operators $\JC_{ca}(\BC^c)$ and $\JC_{cb}(\BC)$ are
both constants which only depend on the channel $\BC$.

For any channel $\BC$, the SDP~\eqref{SDPRepEpsD} can be solved numerically to
compute $\ep_{\BC}$. However, $\ep_{\BC}$ alone is not sufficient to evaluate
the bound~\eqref{approxDeg} on the quantum capacity of $\QC(\BC)$. In addition,
one requires $\QC^{(1)}(\BC)$. As stated earlier, methods for computing
$\QC^{(1)}(\BC)$, obtained from solving a non-convex optimization
problem~\eqref{eq:CICal}, are limited. This limitation makes it non-trivial to
evaluate the bound~\eqref{approxDeg} for an abritrary channel $\BC$. However,
there are special well studied channels for which $\QC^{(1)}(\BC)$ is
known~\cite{DiVincenzoShorEA98, WolfPerezGarcia07, FernWhaley08,
SiddhuGriffiths16}.  These include channels that are degradable and have
$\ep_{\BC} = 0$, but also include channels that are not degradable. 

One simple example of a degradable channel is the qubit dephasing channel with
dephasing probability $q$~\cite{WolfPerezGarcia07},
\begin{equation}
    \FC_q(\rho) = (1-q) \rho + q Z \rho Z^{\dag},
    \label{eq:dephasing}
\end{equation}
where $Z$ is the Pauli matrix defined in~\eqref{pauliMat}. The coherent
information of this channel, $\QC^{(1)}(\FC_q)$, is simply $I_c(\FC_q, I/2) = 1
- h(q)$. Since $\FC_q$ channel is degradable, $\ep_{\FC_q} = 0$.
One may verify this by solving the SDP~\eqref{SDPRepEpsD} to obtain a numerical
value $\ep^*_{\FC_q}$ of $\ep_{\FC_q}$. We find the absolute difference between
$\ep^*_{\FC_q}$ and $\ep_{\FC_q}$ to be small, $O(10^{-10})$. Using
$\ep_{\FC_q}$, an equation of the form~\eqref{approxDeg} for $\FC_q$ gives
$\QC^{(1)}(\FC_q) \leq \QC(\FC_q) \leq \QC^{(1)}(\FC_1)$.  These inequalities
simply state that the quantum capacity of $\FC_q$ equals its coherent
information, $1 - h(q)$.

Another simple example of a degradable channel is the erasure channel. Recall,
the erasure channel $\EC_{p}$ with erasure probability $p$ acting on a
$d_a$-dimensional input is defined in~\eqref{eraChan}. We consider the case
where $0 \leq p \leq 1/2$. For these values of $p$, the erasure channel is
degradable, $\ep_{\EC_p} = 0$ and agrees with its numerical estimate
$\ep^*_{\EC_q}$ up to $O(10^{-9})$. This estimate is obtained by solving the
SDP~\eqref{SDPRepEpsD} for a qubit erasure channel with erasure probability
chosen randomly between zero and one half.
The channel coherent information, $\QC^{(1)}(\EC_{p})$, simply equals
$I_c(\EC_{p}, I/d_a) = (1-2p) \log d_a$~(see Sec.4
in~\cite{SiddhuGriffiths21}). These results, together with~\eqref{approxDeg},
imply that $\QC^{(1)}(\EC_p) = \QC(\EC_{p})$ for $p \leq 1/2$. This equality
also holds for $1/2 < p \leq 1$, where $\EC_p$ is anti-degradability.

As our final example, we consider the qubit depolarizing channel, $\Dl$,
with Kraus decomposition~\eqref{krausOpDepol},
\begin{equation}
    \Dl(\rho) = (1-p) \rho + \frac{p}{3}(X \rho X + Y \rho Y + Z \rho Z ),
    \label{eq:depolKraus}
\end{equation}
where $0 \leq p \leq 1$. The coherent information of this channel is
\begin{equation}
    \QC^{(1)}(\Dl) = \max(0, 1-h(p) - p \log_2 3).
\end{equation}
The qubit depolarizing channel $\Dl$ is not degradable, except at $p=0$.  The
smallest $\ep$ for which equation~\eqref{epsDeg} holds must be found by
numerically solving an SDP~\eqref{SDPRepEpsD}. This numerical solution, along
with~\eqref{approxDeg}, can be used to compute bounds on the quantum
capacity of $\Dl$. 
For instance, when the depolarizing parameter $p \simeq .026$, solving the
SDP~\eqref{SDPRepEpsD} gives a numerical value $\ep_{\DC}^* \simeq 3.8 \times
10^{-3}$ of $\ep_{\DC}$. Using this value, one finds that $|\QC(\Dl) -
\QC^{(1)}(\Dl)| \leq 5.1 \times 10^{-2}$.
For these and other numerics of this type, a short Python based
notebook accompanies chapter~(see Notebook~5 in~\cite{NoName21a}).
This same notebook has additional examples consisting of the qubit dephasing
and erasure channels.

So far we have discussed $\ep$-degradable channels, which is one approximate
notion of degradability. This notion allows one to find computable bounds on
the quantum capacity of channels with known coherent information. Another
notion for approximate degradability is $\ep$-close degradable
channels~\cite{SutterScholzEA17}. A given channel $\BC$ is $\ep$-close
degradable if there is a degradable channel $\MC$ which is $\ep$ close to $\BC$
in diamond norm distance, i.e., $|| \BC - \MC||_{\diamond} = \ep$.  If $\BC$ is
$\ep$-close degradable, then its quantum capacity can be bounded as follows
\begin{equation}
    |\QC(\BC) - \QC^{(1)}(\MC)| \leq \ep \log(d_b) + (2 + \ep) h(\frac{\ep}{2+\ep}).
    \label{epCloseBound}
\end{equation}
In the expression above $\QC^{(1)}(\MC)$ can be computed using tools from
convex optimization because $\MC$ is degradable. On the other hand, there is
no known way to efficiently compute the smallest $\ep$ for which a given
channel $\BC$ is $\ep$-close degradable. Such computations could potentially
lead to new and possibly tighter bounds~\eqref{epCloseBound} on the quantum
capacity of channels. Finding such bounds is an active area of fundamental
research in quantum Shannon theory~\cite{HolevoWerner01, SmithSmolin08,
Ouyang14, WangDuan16, WangFangEA18, FangFawzi21, FanizzaKianvashEA20}. 

\section{Concluding Remarks}
\label{sec:cncl}
\textcolor{black}{The SDP models were chosen here so that they can be accompanied by working numerical examples to aid in learning, as a companion to closed-form solutions possible in special situations.
However, practical use of SDPs for QIS can easily become numerically intractable for
current solvers as the number of quantum systems involved increase (a quantum system consisting of $k$ qubits has dimension $d=2^k$ and it introduces $O(2^k)$ variables, which is exponential in
$k$.) This can make the SDPs in
Secs.~\ref{P1StDis},\ref{P2StFid},\ref{P3ChDis}, and \ref{P4EntSep} numerically
intractable for larger $k$. This is not the only manner in which SDPs in QIS become intractable. As one example, the quantum query complexity of a Boolean function
of $n$-bits can be computed using an SDP that has $O(2^n)$
variables(see~\cite{Reichardt10, Reichardt10a} and references therein). Another reason is non-linearity: certain optimization problems
involving the von-Neumann entropy, a non-linear function, can be approximated
using an SDP~\cite{FawziFawzi18, FawziSaundersonEA19}, with numerical effort increasing with precision, sometimes prohibitively. We hope that this introduction to QIS through these starter problems provides a rapid and accessible gateway to new researchers.}

\section*{Acknowledgments}
VS gratefully acknowledges support from NSF CAREER Award CCF 1652560 and NSF
grant PHY 1915407, thanks Mark M. Wilde for helpful comments, 
and thanks to Maximilian Stahlberg for support with PICOS.

\appendix

    \section{Norms of Superoperators}
\label{AppNorm}
In Sec.~\ref{P3ChDis} we defined a superoperator $\DC$~\eqref{eq:del}, which
can we written as,
\begin{equation}
    \DC (\rho) = \frac{(1-\lm)}{2} (\IC(\rho) - \TC(\rho)),
    \label{eq:ADel}
\end{equation}
where $\IC(\rho) = \rho$ and $\TC(\rho) = \Tr(\rho) \frac{I}{d}$.  Here we compute the
$l_1$-norm and the diamond norm of this superoperator.
The $l_1$-norm of $\DC$ is given by~\eqref{prodMax2}
\begin{equation}
    ||\DC||_1 = \max_{\rho} ||\DC(\rho)||_1.
    \label{eq:ADelN1}
\end{equation}
The set of density operators is convex and $||.||_1$, the $1-$norm of an operator, is a
convex function. As a result, the optimum value of the convex maximization problem
above is achieved at an extreme point of the set of density operator. These extreme
points are projectors onto pure states, i.e., $\dya{\psi}$ where $\inp{\psi} = 1$.
Thus
\begin{equation}
    ||\DC||_1 = \max_{\ket{\psi}} ||\DC(\dya{\psi})||_1.
    \label{eq:ADelN1B}
\end{equation}
A simple calculation shows that for $\DC$ in~\eqref{eq:ADel}, $\DC(\dya{\psi})$
is independent of $\ket{\psi}$ and equals $(1-\lm)(d-1)/d$. 

To compute the diamond norm of $\DC$ in~\eqref{eq:ADel} we use a technique similar
to the one employed in~\cite{LeditzkyLeungEA18a} to compute $||\DC||_{\diamond}$ for $d=2$.
Using~\eqref{entMax}, we can bound $||\DC||_{\diamond}$ from below by
$||\IC_d \ot \DC(X)||_1$, where $||X||_1 \leq 1$. Letting $X = \dya{\phi_d}$,
where
\begin{equation}
    \ket{\phi_d} = \frac{1}{\sqrt{d}} \sum_{i=1}^d \ket{i} \ot \ket{i}
\end{equation}
is the maximally entangled state on two qudits, gives
\begin{equation}
    ||\DC||_{\diamond} \geq ||\DC(\dya{\phi}) \ot \IC_d ||_1 = (1-\lm) \frac{d^2 - 1}{d^2}.
    \label{eq:ALB}
\end{equation}
This lower bound can be matched by an upper bound. To obtain this upper bound
notice $\DC$ in~\eqref{eq:ADel} is proportional to $\NC = \IC - \TC$, the
difference of two quantum channels. The diamond norm of such a difference is
given by the SDP~\cite{Watrous09}:
\begin{align}
    \begin{aligned}
        \text{minimize} \; & 2 \mu &\\
        \text{subject to} \;& \mu I_a \succeq \Tr_b(Z_{ba}), &\\
        &Z_{ba} \succeq \JC_{ba}(\NC), &\\ 
        \text{and}\; & Z_{ab} \succeq 0.
    \end{aligned}
\end{align}
Notice $Z_{ba} = \frac{d^2 - 1}{d} \dya{\phi_d}$ is a feasible solution of this SDP
with $\mu = \frac{d^2 - 1}{d^2}$, thus $||\NC||_{\diamond} \leq
2\frac{d^2 - 1}{d^2}$.  Since $||\DC||_{\diamond} =
\frac{(1-\lm)}{2}||N||_{\diamond}$, we get a matching upper bound
to~\eqref{eq:ALB}.

\section{Transfer Matrix and Choi-Jamio\l{}kowski representation}
\label{AppLin}

Let $\HC$ be a $d$-dimensional complex space and $\LC(\HC)$ be the space of linear
operators on $\HC$. Given two operators $A$ and $B$ in $\LC(\HC)$, their Frobenius 
inner product is
\begin{equation}
    \inpV{A}{B} := \Tr(A^{\dag} B).
\end{equation}
Using this inner product, one can define an orthonormal basis of $d^2$
operators, $\{ \dyad{i}{j}\}$, for $\LC(\HC)$.  Using such a basis, a channel
superoperator $\BC:\LC(\HC_a) \to \LC(\HC_b)$ can be written in matrix
form,
\begin{equation}
    \BC(\dyad{k}{l}_a) = \sum_{i,j} T(\BC)_{kl,ij} \dyad{i}{j}_b,
    \label{ap:txfrMt}
\end{equation}
where the complex numbers
\begin{equation}
    T(\BC)_{kl,ij} = \inpV{\BC(\dyad{k}{l}_a)}{\dyad{i}{j}_b},
    \label{ap:txfrMtE}
\end{equation}
form the entries of a $d_b^2 \times d_a^2$ matrix $T(\BC)$, sometimes called
the {\em transfer matrix} of the channel $\BC$. 
The rows and columns of $T(\BC)$ are indexed by $0 \leq k,l \leq d_b-1$ and $0
\leq i,j \leq d_a -1$ respectively. This transfer matrix $T(\BC) : \HC_{a} \ot \HC_a
\to \HC_{b} \ot \HC_b$ is related to the channel's Choi-Jamio\l{}kowski
representation $\JC_{ba}(\BC) \in \HC_{ba}$~(defined in~\eqref{ChoiJ}) as
follows
\begin{equation}
    \bra{kl} T(\BC) \ket{ij} = \bra{ki} \JC_{ba}(\BC) \ket{lj},
    \label{ap:TxfrCJ}
\end{equation}
where we use the notation $\ket{ij}:= \ket{i} \ot \ket{j}$.

Let $\BC: \LC(\HC_a) \to \LC(\HC_b)$ and $\CC: \LC(\HC_b) \to
\LC(\HC_c)$ be two channels with transfer matrices $T(\CC)$ and $T(\BC)$,
respectively. Using the two channels in series results in a third channel $\CC
\circ \BC$ with transfer matrix $T(\CC \circ \BC)$. This matrix is simply the
product of the individual transfer matrices and is given by
\begin{equation}
    T(\CC \circ \BC) = T(\CC) T(\BC),
    \label{ap:serTxf}
\end{equation}
Using the above equation, along with the relation~\eqref{ap:TxfrCJ} one can
show that $\JC_{ca}(\CC \circ \BC)$, the Choi-Jamio\l{}kowski representation of
$\CC \circ \BC$, is just a linear function of $\JC_{ba}(\BC)$ and
$\JC_{cb}(\CC)$.


\begin{thebibliography}{100}
\expandafter\ifx\csname url\endcsname\relax
  \def\url#1{\texttt{#1}}\fi
\expandafter\ifx\csname doi\endcsname\relax
  \def\doi#1{\burlalt{doi:#1}{http://dx.doi.org/#1}}\fi
\expandafter\ifx\csname urlprefix\endcsname\relax\def\urlprefix{}\fi
\expandafter\ifx\csname href\endcsname\relax
  \def\href#1#2{#2}\fi
\expandafter\ifx\csname burlalt\endcsname\relax
  \def\burlalt#1#2{\href{#2}{#1}}\fi

\bibitem{NielsenChuang11}
Michael~A. Nielsen and Isaac~L. Chuang.
\newblock {\em Quantum Computation and Quantum Information: 10th Anniversary
  Edition}.
\newblock Cambridge University Press, New York, NY, USA, 10th edition, 2011.

\bibitem{Shor97}
Peter~W. Shor.
\newblock Polynomial-time algorithms for prime factorization and discrete
  logarithms on a quantum computer.
\newblock {\em SIAM Journal on Computing}, 26(5):1484–1509, Oct 1997.
\newblock \doi{10.1137/s0097539795293172}.

\bibitem{HorodeckiHorodeckiEA09a}
Ryszard Horodecki, Pawe\l{} Horodecki, Micha\l{} Horodecki, and Karol
  Horodecki.
\newblock Quantum entanglement.
\newblock {\em Rev. Mod. Phys.}, 81:865--942, Jun 2009.
\newblock \doi{10.1103/RevModPhys.81.865}.

\bibitem{BennettBrassard84}
C.~H. Bennett and G.~Brassard.
\newblock Quantum cryptography: Public key distribution and coin tossing.
\newblock In {\em Proceedings of IEEE International Conference on Computers,
  Systems, and Signal Processing}, page 175, India, 1984.

\bibitem{BennettShor98}
C.~H. Bennett and P.~W. Shor.
\newblock Quantum information theory.
\newblock {\em IEEE Transactions on Information Theory}, 44(6):2724--2742, Oct
  1998.
\newblock \doi{10.1109/18.720553}.

\bibitem{SmithYard08}
Graeme Smith and Jon Yard.
\newblock Quantum communication with zero-capacity channels.
\newblock {\em Science}, 321(5897):1812--1815, 2008.
\newblock \doi{10.1126/science.1162242}.

\bibitem{BennettShor04}
Charles~H. Bennett and Peter~W. Shor.
\newblock Quantum channel capacities.
\newblock {\em Science}, 303(5665):1784--1787, 2004,
  \burlalt{https://www.science.org/doi/pdf/10.1126/science.1092381}{http://arxiv.org/abs/https://www.science.org/doi/pdf/10.1126/science.1092381}.
\newblock \doi{10.1126/science.1092381}.

\bibitem{VandenbergheBoyd96}
Lieven Vandenberghe and Stephen Boyd.
\newblock Semidefinite programming.
\newblock {\em SIAM Review}, 38(1):49--95, 1996.
\newblock \doi{10.1137/1038003}.

\bibitem{OvertonWolkowicz97}
Michael Overton and Henry Wolkowicz.
\newblock Semidefinite programming.
\newblock {\em Mathematical Programming}, 77(1):105--109, Apr 1997.
\newblock \doi{10.1007/BF02614431}.

\bibitem{Todd01}
M.~J. Todd.
\newblock Semidefinite optimization.
\newblock {\em Acta Numerica}, 10:515–560, 2001.
\newblock \doi{10.1017/S0962492901000071}.

\bibitem{BellmanFan63}
Richard Bellman and Ky~Fan.
\newblock On systems of linear inequalities in hermitian matrix variables.
\newblock {\em Proc. Sympos. Pure Math.}, VII:1--11, 1963.

\bibitem{DantzigThapa97}
George~B. Dantzig and Mukund~N. Thapa.
\newblock {\em Linear Programming 1}.
\newblock Springer-Verlag New York, 1997.
\newblock \doi{10.1007/b97672}.

\bibitem{Schrijver98}
Alexander Schrijver.
\newblock {\em Theory of Linear and Integer Programming}.
\newblock John Wiley \& Sons, 1998.

\bibitem{Pataki96}
G{\'a}bor Pataki.
\newblock Cone-lp's and semidefinite programs: Geometry and a simplex-type
  method.
\newblock In William~H. Cunningham, S.~Thomas McCormick, and Maurice Queyranne,
  editors, {\em Integer Programming and Combinatorial Optimization}, pages
  162--174, Berlin, Heidelberg, 1996. Springer Berlin Heidelberg.

\bibitem{BoydElGhaouiEA94}
Stephen Boyd, Laurent El~Ghaoui, Eric Feron, and Venkataramanan Balakrishnan.
\newblock {\em Linear Matrix Inequalities in System and Control Theory}.
\newblock Society for Industrial and Applied Mathematics, 1994.
\newblock \doi{10.1137/1.9781611970777}.

\bibitem{NemirovskyYudin77}
A.S. Nemirovsky and D.B. Yudin.
\newblock Informational complexity and efficient methods for solving complex
  extremal problems.
\newblock {\em Matekon}, 13:25--45, 1977.

\bibitem{Shor77}
N.~Z. Shor.
\newblock Cut-off method with space extension in convex programming problems.
\newblock {\em Cybernetics}, 13(1):94--96, Jan 1977.
\newblock \doi{10.1007/BF01071394}.

\bibitem{Alizadeh92}
Farid Alizadeh.
\newblock {\em Combinatorial Optimization with Interior Point Methods and
  Semi-Definite Matrices}.
\newblock PhD thesis, University of Minnesota, USA, 1992.
\newblock UMI Order No. GAX92-07776.

\bibitem{NesterovNemirovskii94}
Yurii Nesterov and Arkadii Nemirovskii.
\newblock {\em Interior-Point Polynomial Algorithms in Convex Programming}.
\newblock Society for Industrial and Applied Mathematics, 1994.
\newblock \doi{10.1137/1.9781611970791}.

\bibitem{KamathKarmarkar14}
A.~Kamath and N.~Karmarkar.
\newblock {\em A Continuous Approach to Compute Upper Bounds in Quadratic
  Maximization Problems With Integer Constraints:}, pages 125--140.
\newblock Princeton University Press, 2014.
\newblock \doi{10.1515/9781400862528.125}.

\bibitem{Karmarkar84}
N.~Karmarkar.
\newblock A new polynomial-time algorithm for linear programming.
\newblock {\em Combinatorica}, 4(4):373--395, Dec 1984.
\newblock \doi{10.1007/BF02579150}.

\bibitem{GoemansWilliamson95}
Michel~X. Goemans and David~P. Williamson.
\newblock Improved approximation algorithms for maximum cut and satisfiability
  problems using semidefinite programming.
\newblock {\em J. ACM}, 42(6):1115–1145, November 1995.
\newblock \doi{10.1145/227683.227684}.

\bibitem{KargerMotwaniEA98}
David Karger, Rajeev Motwani, and Madhu Sudan.
\newblock Approximate graph coloring by semidefinite programming.
\newblock {\em J. ACM}, 45(2):246–265, March 1998.
\newblock \doi{10.1145/274787.274791}.

\bibitem{Alizadeh95}
Farid Alizadeh.
\newblock Interior point methods in semidefinite programming with applications
  to combinatorial optimization.
\newblock {\em SIAM Journal on Optimization}, 5(1):13--51, 1995.
\newblock \doi{10.1137/0805002}.

\bibitem{HauensteinLiddellEA21}
Jonathan~D. Hauenstein, Alan~C. Liddell, Sanesha McPherson, and Yi~Zhang.
\newblock Numerical algebraic geometry and semidefinite programming.
\newblock {\em Results in Applied Mathematics}, 11:100166, 2021.
\newblock \doi{10.1016/j.rinam.2021.100166}.

\bibitem{Watrous18}
John Watrous.
\newblock {\em The {Theory} of {Quantum} {Information}}.
\newblock Cambridge University Press, 1 edition, Apr 2018.
\newblock \doi{10.1017/9781316848142}.

\bibitem{Wang18}
Xin Wang.
\newblock {\em Semidefinite optimization for quantum information}.
\newblock PhD thesis, University of Technology Sydney, Jul 2018.
\newblock \urlprefix\url{http://hdl.handle.net/10453/127996}.

\bibitem{KhatriWilde20}
Sumeet Khatri and Mark~M. Wilde.
\newblock Principles of quantum communication theory: A modern approach, 2020.
\newblock \burlalt{2011.04672}{http://arxiv.org/abs/2011.04672}.

\bibitem{CoverThomas01}
Thomas~M. Cover and Joy~A. Thomas.
\newblock {\em Elements of Information Theory}.
\newblock John Wiley \& Sons, Ltd, 2001.
\newblock \doi{10.1002/0471200611}.

\bibitem{Schumacher95}
Benjamin Schumacher.
\newblock Quantum coding.
\newblock {\em Phys. Rev. A}, 51:2738--2747, Apr 1995.
\newblock \doi{10.1103/PhysRevA.51.2738}.

\bibitem{SudarshanMathewsEA61}
E.~C.~G. Sudarshan, P.~M. Mathews, and Jayaseetha Rau.
\newblock Stochastic dynamics of quantum-mechanical systems.
\newblock {\em Phys. Rev.}, 121:920--924, Feb 1961.
\newblock \doi{10.1103/PhysRev.121.920}.

\bibitem{Holevo12}
A.~S. Holevo.
\newblock {\em Quantum Systems, Channels, Information}.
\newblock De Gruyter, 2012.
\newblock \doi{doi:10.1515/9783110273403.346}.

\bibitem{Wilde17}
Mark~M. Wilde.
\newblock {\em Quantum Information Theory}.
\newblock Cambridge University Press, 2 edition, 2017.
\newblock \doi{10.1017/9781316809976}.

\bibitem{WolfPerezGarcia07}
Michael~M. Wolf and David {P\'erez-Garc\'\i a}.
\newblock Quantum capacities of channels with small environment.
\newblock {\em Phys. Rev. A}, 75:012303, Jan 2007.
\newblock \doi{10.1103/PhysRevA.75.012303}.

\bibitem{KhatriSharmaEA20}
Sumeet Khatri, Kunal Sharma, and Mark~M. Wilde.
\newblock Information-theoretic aspects of the generalized amplitude-damping
  channel.
\newblock {\em Phys. Rev. A}, 102:012401, Jul 2020.
\newblock \doi{10.1103/PhysRevA.102.012401}.

\bibitem{HolevoGiovannetti12}
A~S Holevo and V~Giovannetti.
\newblock Quantum channels and their entropic characteristics.
\newblock {\em Reports on Progress in Physics}, 75(4):046001, 2012.
\newblock \urlprefix\url{http://stacks.iop.org/0034-4885/75/i=4/a=046001}.

\bibitem{Kraus83}
K.~Kraus.
\newblock {\em States, Effects, and Operations Fundamental Notions of Quantum
  Theory}.
\newblock Springer, Berlin, Heidelberg, 1983.
\newblock \doi{10.1007/3-540-12732-1}.

\bibitem{Choi75}
Man-Duen Choi.
\newblock Completely positive linear maps on complex matrices.
\newblock {\em Linear Algebra and its Applications}, 10(3):285 -- 290, 1975.
\newblock \doi{10.1016/0024-3795(75)90075-0}.

\bibitem{Jamiokowski72}
A.~Jamio\l{}kowski.
\newblock Linear transformations which preserve trace and positive
  semidefiniteness of operators.
\newblock {\em Reports on Mathematical Physics}, 3(4):275--278, 1972.
\newblock \doi{10.1016/0034-4877(72)90011-0}.

\bibitem{HorodeckiShorEA03}
Michael Horodecki, Peter~W. Shor, and Mary~Beth Ruskai.
\newblock Entanglement breaking channels.
\newblock {\em Reviews in Mathematical Physics}, 15(06):629--641, 2003.
\newblock \doi{10.1142/S0129055X03001709}.

\bibitem{Helstrom69}
Carl~W. Helstrom.
\newblock Quantum detection and estimation theory.
\newblock {\em Journal of Statistical Physics}, 1(2):231--252, Jun 1969.
\newblock \doi{10.1007/BF01007479}.

\bibitem{SagnolStahlberg21}
G.~Sagnol and M.~Stahlberg.
\newblock Picos: A python interface to conic optimization solvers, Dec 2021.
\newblock \urlprefix\url{https://cvxopt.org/index.html}.

\bibitem{AndersenDahlEA21}
Martin {Andersen}, Joachim {Dahl}, and Lieven {Vandenberghe}.
\newblock Cvxopt: Convex optimization, Sep 2021.
\newblock \urlprefix\url{https://cvxopt.org/index.html}.

\bibitem{NoName21a}
Vikesh Siddhu and Sridhar Tayur.
\newblock Sdp-quantum-or.
\newblock \url{https://github.com/vsiddhu/SDP-Quantum-OR}, 2021.

\bibitem{Kholevo79}
A.~S. Kholevo.
\newblock On asymptotically optimal hypothesis testing in quantum statistics.
\newblock {\em Theory of Probability \& Its Applications}, 23(2):411--415,
  1979.
\newblock \doi{10.1137/1123048}.

\bibitem{Kargin05}
Vladislav Kargin.
\newblock On the chernoff bound for efficiency of quantum hypothesis testing.
\newblock {\em The Annals of Statistics}, 33(2):959 -- 976, 2005.
\newblock \doi{10.1214/009053604000001219}.

\bibitem{AudenaertCalsamigliaEA07}
K.~M.~R. Audenaert, J.~Calsamiglia, R.~Mu\~noz Tapia, E.~Bagan, Ll. Masanes,
  A.~Acin, and F.~Verstraete.
\newblock Discriminating states: The quantum chernoff bound.
\newblock {\em Phys. Rev. Lett.}, 98:160501, Apr 2007.
\newblock \doi{10.1103/PhysRevLett.98.160501}.

\bibitem{Hayashi17}
Masahito Hayashi.
\newblock {\em Quantum Hypothesis Testing and Discrimination of Quantum
  States}, pages 95--153.
\newblock Springer Berlin Heidelberg, Berlin, Heidelberg, 2017.
\newblock \doi{10.1007/978-3-662-49725-8\_3}.

\bibitem{NussbaumSzkoa09}
Michael Nussbaum and Arleta Szko\l{}a.
\newblock The chernoff lower bound for symmetric quantum hypothesis testing.
\newblock {\em The Annals of Statistics}, 37(2):1040 -- 1057, 2009.
\newblock \doi{10.1214/08-AOS593}.

\bibitem{Parthasarathy01}
K.~R. Parthasarathy.
\newblock {\em On Consistency of the Maximum Likelihood Method in Testing
  Multiple Quantum Hypotheses}, pages 361--377.
\newblock Birkh{\"a}user Boston, Boston, MA, 2001.
\newblock \doi{10.1007/978-1-4612-0167-0\_19}.

\bibitem{Chefles00a}
Anthony Chefles.
\newblock Quantum state discrimination.
\newblock {\em Contemporary Physics}, 41(6):401--424, 2000.
\newblock \doi{10.1080/00107510010002599}.

\bibitem{BarnettCroke09}
Stephen~M. Barnett and Sarah Croke.
\newblock Quantum state discrimination.
\newblock {\em Adv. Opt. Photon.}, 1(2):238--278, Apr 2009.
\newblock \doi{10.1364/AOP.1.000238}.

\bibitem{BaeKwek15}
Joonwoo Bae and Leong-Chuan Kwek.
\newblock Quantum state discrimination and its applications.
\newblock {\em Journal of Physics A: Mathematical and Theoretical},
  48(8):083001, Jan 2015.
\newblock \doi{10.1088/1751-8113/48/8/083001}.

\bibitem{Uhlmann76}
A.~Uhlmann.
\newblock The "transition probability" in the state space of a *-algebra.
\newblock {\em Reports on Mathematical Physics}, 9(2):273--279, 1976.
\newblock \doi{10.1016/0034-4877(76)90060-4}.

\bibitem{Bhattacharya43}
Anil~K. Bhattacharya.
\newblock On a measure of divergence between two statistical populations
  defined by their probability distributions.
\newblock {\em Bulletin of Calcutta Mathematical Society}, 35:99, 1943.

\bibitem{Killoran12}
Nathan Killoran.
\newblock {\em Entanglement quantification and quantum benchmarking of optical
  communication devices}.
\newblock PhD thesis, University of Waterloo, Apr 2012.
\newblock \urlprefix\url{http://hdl.handle.net/10012/6662}.

\bibitem{Watrous12}
John Watrous.
\newblock Simpler semidefinite programs for completely bounded norms.
\newblock {\em Chicago Journal of Theoretical Computer Science}, 19, 07 2012.
\newblock \doi{10.4086/cjtcs.2013.008}.

\bibitem{ZhangChenEA15}
Lin Zhang, Lin Chen, and Kaifeng Bu.
\newblock Fidelity between one bipartite quantum state and another undergoing
  local unitary dynamics.
\newblock {\em Quantum Information Processing}, 14(12):4715--4730, Dec 2015.
\newblock \doi{10.1007/s11128-015-1117-7}.

\bibitem{YuanFung17}
Haidong Yuan and Chi-Hang~Fred Fung.
\newblock Fidelity and fisher information on quantum channels.
\newblock {\em New Journal of Physics}, 19(11):113039, Nov 2017.
\newblock \doi{10.1088/1367-2630/aa874c}.

\bibitem{GutoskiRosmanisEA18}
Gus Gutoski, Ansis Rosmanis, and Jamie Sikora.
\newblock Fidelity of quantum strategies with applications to cryptography.
\newblock {\em Quantum}, 2:89, September 2018.
\newblock \doi{10.22331/q-2018-09-03-89}.

\bibitem{KatariyaWilde21}
Vishal Katariya and Mark~M. Wilde.
\newblock Geometric distinguishability measures limit quantum channel
  estimation and discrimination.
\newblock {\em Quantum Information Processing}, 20(2):78, Feb 2021.
\newblock \doi{10.1007/s11128-021-02992-7}.

\bibitem{Watrous09}
John Watrous.
\newblock Semidefinite programs for completely bounded norms.
\newblock {\em Theory of Computing}, 5(11):217--238, 2009.
\newblock \doi{10.4086/toc.2009.v005a011}.

\bibitem{Watrous08}
John Watrous.
\newblock Distinguishing quantum operations having few kraus operators.
\newblock {\em Quantum Info. Comput.}, 8(8):819–833, September 2008.

\bibitem{HolevoWerner01}
Alexander~S Holevo and Reinhard~F Werner.
\newblock Evaluating capacities of bosonic gaussian channels.
\newblock {\em Physical Review A}, 63(3):032312, 2001,
  \burlalt{quant-ph/9912067}{http://arxiv.org/abs/quant-ph/9912067}.

\bibitem{WangDuan16}
Xin Wang and Runyao Duan.
\newblock A semidefinite programming upper bound of quantum capacity.
\newblock In {\em 2016 IEEE International Symposium on Information Theory
  (ISIT)}, pages 1690--1694, 2016.
\newblock \doi{10.1109/ISIT.2016.7541587}.

\bibitem{WangWilde18}
Xin Wang and Mark~M. Wilde.
\newblock Exact entanglement cost of quantum states and channels under
  ppt-preserving operations, 2018.
\newblock \burlalt{1809.09592}{http://arxiv.org/abs/1809.09592}.

\bibitem{IyerPoulin17}
Pavithran~S. Iyer and David Poulin.
\newblock A small quantum computer is needed to optimize fault-tolerant
  protocols, 2017.
\newblock \burlalt{1711.04736}{http://arxiv.org/abs/1711.04736}.

\bibitem{MagesanGambettaEA12}
Easwar Magesan, Jay~M. Gambetta, and Joseph Emerson.
\newblock Characterizing quantum gates via randomized benchmarking.
\newblock {\em Phys. Rev. A}, 85:042311, Apr 2012.
\newblock \doi{10.1103/PhysRevA.85.042311}.

\bibitem{KlieschKuengEA16}
Martin Kliesch, Richard Kueng, Jens Eisert, and David Gross.
\newblock Improving compressed sensing with the diamond norm.
\newblock {\em IEEE Transactions on Information Theory}, 62(12):7445--7463,
  2016.
\newblock \doi{10.1109/TIT.2016.2606500}.

\bibitem{FannesLewisEA88}
M.~Fannes, J.~T. Lewis, and A.~Verbeure.
\newblock Symmetric states of composite systems.
\newblock {\em Letters in Mathematical Physics}, 15(3):255--260, Apr 1988.
\newblock \doi{10.1007/BF00398595}.

\bibitem{RaggioWerner89}
G.~Raggio and Reinhard Werner.
\newblock Quantum statistical mechanics of general mean field systems.
\newblock {\em Helvetica Physica Acta}, 62:980, 03 1989.
\newblock \doi{10.5169/seals-116175}.

\bibitem{Werner89}
Reinhard~F. Werner.
\newblock An application of bell's inequalities to a quantum state extension
  problem.
\newblock {\em Letters in Mathematical Physics}, 17(4):359--363, May 1989.
\newblock \doi{10.1007/BF00399761}.

\bibitem{Peres96}
Asher Peres.
\newblock Separability criterion for density matrices.
\newblock {\em Phys. Rev. Lett.}, 77:1413--1415, Aug 1996.
\newblock \doi{10.1103/PhysRevLett.77.1413}.

\bibitem{HorodeckiHorodeckiEA96}
Micha\l{} Horodecki, Pawe\l{} Horodecki, and Ryszard Horodecki.
\newblock Separability of mixed states: necessary and sufficient conditions.
\newblock {\em Physics Letters A}, 223(1):1--8, 1996.
\newblock \doi{10.1016/S0375-9601(96)00706-2}.

\bibitem{DohertyParriloEA04}
Andrew~C. Doherty, Pablo~A. Parrilo, and Federico~M. Spedalieri.
\newblock Complete family of separability criteria.
\newblock {\em Physical Review A}, 69(2), Feb 2004.
\newblock \doi{10.1103/physreva.69.022308}.

\bibitem{HorodeckiHorodeckiEA01}
Michal Horodecki, Pawel Horodecki, and Ryszard Horodecki.
\newblock Mixed-state entanglement and quantum communication, 2001.
\newblock \burlalt{quant-ph/0109124}{http://arxiv.org/abs/quant-ph/0109124}.

\bibitem{NavascuesPironioEA08}
Miguel Navascu{\'{e}}s, Stefano Pironio, and Antonio Ac{\'{\i}}n.
\newblock A convergent hierarchy of semidefinite programs characterizing the
  set of quantum correlations.
\newblock {\em New Journal of Physics}, 10(7):073013, jul 2008.
\newblock \doi{10.1088/1367-2630/10/7/073013}.

\bibitem{BertaFawziEA16}
Mario Berta, Omar Fawzi, and Volkher~B. Scholz.
\newblock Quantum bilinear optimization.
\newblock {\em SIAM Journal on Optimization}, 26(3):1529–1564, Jan 2016.
\newblock \doi{10.1137/15m1037731}.

\bibitem{HarrowNatarajanEA17}
Aram~W. Harrow, Anand Natarajan, and Xiaodi Wu.
\newblock An improved semidefinite programming hierarchy for testing
  entanglement.
\newblock {\em Communications in Mathematical Physics}, 352(3):881–904, Mar
  2017.
\newblock \doi{10.1007/s00220-017-2859-0}.

\bibitem{HarrowNatarajanEA19}
Aram~W. Harrow, Anand Natarajan, and Xiaodi Wu.
\newblock Limitations of semidefinite programs for separable states and
  entangled games.
\newblock {\em Communications in Mathematical Physics}, 366(2):423–468, Mar
  2019.
\newblock \doi{10.1007/s00220-019-03382-y}.

\bibitem{Shannon48}
C.~E. Shannon.
\newblock A mathematical theory of communication.
\newblock {\em Bell System Technical Journal}, 27(3):379--423, 1948.
\newblock \doi{10.1002/j.1538-7305.1948.tb01338.x}.

\bibitem{Blahut72}
R.~Blahut.
\newblock Computation of channel capacity and rate-distortion functions.
\newblock {\em IEEE Transactions on Information Theory}, 18(4):460--473, 1972.
\newblock \doi{10.1109/TIT.1972.1054855}.

\bibitem{Arimoto72}
S.~Arimoto.
\newblock An algorithm for computing the capacity of arbitrary discrete
  memoryless channels.
\newblock {\em IEEE Transactions on Information Theory}, 18(1):14--20, 1972.
\newblock \doi{10.1109/TIT.1972.1054753}.

\bibitem{BarnumNielsenEA98}
Howard Barnum, M.~A. Nielsen, and Benjamin Schumacher.
\newblock Information transmission through a noisy quantum channel.
\newblock {\em Phys. Rev. A}, 57:4153--4175, Jun 1998.
\newblock \doi{10.1103/PhysRevA.57.4153}.

\bibitem{DiVincenzoShorEA98}
David~P. DiVincenzo, Peter~W. Shor, and John~A. Smolin.
\newblock Quantum-channel capacity of very noisy channels.
\newblock {\em Phys. Rev. A}, 57:830--839, Feb 1998.
\newblock \doi{10.1103/PhysRevA.57.830}.

\bibitem{SmithSmolin07}
Graeme Smith and John~A. Smolin.
\newblock Degenerate quantum codes for pauli channels.
\newblock {\em Phys. Rev. Lett.}, 98:030501, Jan 2007.
\newblock \doi{10.1103/PhysRevLett.98.030501}.

\bibitem{LeditzkyLeungEA18}
Felix Leditzky, Debbie Leung, and Graeme Smith.
\newblock Dephrasure channel and superadditivity of coherent information.
\newblock {\em Phys. Rev. Lett.}, 121:160501, Oct 2018.
\newblock \doi{10.1103/PhysRevLett.121.160501}.

\bibitem{BauschLeditzky20}
Johannes Bausch and Felix Leditzky.
\newblock Quantum codes from neural networks.
\newblock {\em New Journal of Physics}, 22(2):023005, Feb 2020.
\newblock \doi{10.1088/1367-2630/ab6cdd}.

\bibitem{Siddhu20}
Vikesh Siddhu.
\newblock Leaking information to gain entanglement.
\newblock {\em arXiv}, Nov 2020,
  \burlalt{arXiv:2011.15116}{http://arxiv.org/abs/arXiv:2011.15116}.

\bibitem{BauschLeditzky21}
Johannes Bausch and Felix Leditzky.
\newblock Error thresholds for arbitrary pauli noise.
\newblock {\em SIAM Journal on Computing}, 50(4):1410–1460, Jan 2021.
\newblock \doi{10.1137/20m1337375}.

\bibitem{Siddhu21}
Vikesh Siddhu.
\newblock Entropic singularities give rise to quantum transmission.
\newblock {\em Nature Communications}, 12(1):5750, Oct 2021,
  \burlalt{2003.10367}{http://arxiv.org/abs/2003.10367}.
\newblock \doi{10.1038/s41467-021-25954-0}.

\bibitem{Lloyd97}
Seth Lloyd.
\newblock Capacity of the noisy quantum channel.
\newblock {\em Phys. Rev. A}, 55:1613--1622, Mar 1997.
\newblock \doi{10.1103/PhysRevA.55.1613}.

\bibitem{Shor02a}
Peter~W. Shor.
\newblock Quantum error correction, Nov 2002.
\newblock \urlprefix\url{http://www.msri.org/workshops/203/schedules/1181}.

\bibitem{Devetak05}
I.~Devetak.
\newblock The private classical capacity and quantum capacity of a quantum
  channel.
\newblock {\em IEEE Transactions on Information Theory}, 51(1):44--55, Jan
  2005.
\newblock \doi{10.1109/TIT.2004.839515}.

\bibitem{DevetakShor05}
I.~Devetak and P.~W. Shor.
\newblock The capacity of a quantum channel for simultaneous transmission of
  classical and quantum information.
\newblock {\em Communications in Mathematical Physics}, 256(2):287--303, 2005.
\newblock \doi{10.1007/s00220-005-1317-6}.

\bibitem{LeditzkyDattaEA18}
F.~{Leditzky}, N.~{Datta}, and G.~{Smith}.
\newblock Useful states and entanglement distillation.
\newblock {\em IEEE Transactions on Information Theory}, 64(7):4689--4708, July
  2018.
\newblock \doi{10.1109/TIT.2017.2776907}.

\bibitem{YardHaydenEA08}
J.~Yard, P.~Hayden, and I.~Devetak.
\newblock Capacity theorems for quantum multiple-access channels:
  classical-quantum and quantum-quantum capacity regions.
\newblock {\em Information Theory, IEEE Transactions on}, 54(7):3091--3113, Jul
  2008.
\newblock \doi{10.1109/TIT.2008.924665}.

\bibitem{FawziFawzi18}
Hamza Fawzi and Omar Fawzi.
\newblock Efficient optimization of the quantum relative entropy.
\newblock {\em Journal of Physics A: Mathematical and Theoretical},
  51(15):154003, Mar 2018.
\newblock \doi{10.1088/1751-8121/aab285}.

\bibitem{RamakrishnanItenEA21}
Navneeth Ramakrishnan, Raban Iten, Volkher~B. Scholz, and Mario Berta.
\newblock Computing quantum channel capacities.
\newblock {\em IEEE Transactions on Information Theory}, 67(2):946--960, 2021.
\newblock \doi{10.1109/TIT.2020.3034471}.

\bibitem{SutterScholzEA17}
D.~Sutter, V.~B. Scholz, A.~Winter, and R.~Renner.
\newblock Approximate degradable quantum channels.
\newblock {\em IEEE Transactions on Information Theory}, 63(12):7832--7844, Dec
  2017.
\newblock \doi{10.1109/TIT.2017.2754268}.

\bibitem{FernWhaley08}
Jesse Fern and K.~Birgitta Whaley.
\newblock Lower bounds on the nonzero capacity of pauli channels.
\newblock {\em Phys. Rev. A}, 78:062335, Dec 2008.
\newblock \doi{10.1103/PhysRevA.78.062335}.

\bibitem{SiddhuGriffiths16}
Vikesh Siddhu and Robert~B. Griffiths.
\newblock Degradable quantum channels using pure-state to product-of-pure-state
  isometries.
\newblock {\em Phys. Rev. A}, 94:052331, Nov 2016.
\newblock \doi{10.1103/PhysRevA.94.052331}.

\bibitem{SiddhuGriffiths21}
Vikesh Siddhu and Robert~B. Griffiths.
\newblock Positivity and nonadditivity of quantum capacities using generalized
  erasure channels.
\newblock {\em IEEE Transactions on Information Theory}, 67(7):4533--4545,
  2021.
\newblock \doi{10.1109/TIT.2021.3080819}.

\bibitem{SmithSmolin08}
G.~Smith and J.A. Smolin.
\newblock Additive extensions of a quantum channel.
\newblock In {\em Information Theory Workshop, 2008. ITW '08. IEEE}, pages
  368--372, May 2008.
\newblock \doi{10.1109/ITW.2008.4578688}.

\bibitem{Ouyang14}
Yingkai Ouyang.
\newblock Channel covariance, twirling, contraction and some upper bounds on
  the quantum capacity.
\newblock {\em Quantum Information and Computation}, 14:0917--0936, Sept 2014.
\newblock
  \urlprefix\url{http://www.rintonpress.com/xxqic14/qic-14-1112/0917-0936.pdf}.

\bibitem{WangFangEA18}
Xin Wang, Kun Fang, and Runyao Duan.
\newblock Semidefinite programming converse bounds for quantum communication.
\newblock {\em IEEE Transactions on Information Theory}, 65(4):2583--2592,
  2018, \burlalt{1709.00200}{http://arxiv.org/abs/1709.00200}.

\bibitem{FangFawzi21}
Kun Fang and Hamza Fawzi.
\newblock Geometric r{\'e}nyi divergence and its applications in quantum
  channel capacities.
\newblock {\em Communications in Mathematical Physics}, pages 1--63, 2021,
  \burlalt{1909.05758}{http://arxiv.org/abs/1909.05758}.

\bibitem{FanizzaKianvashEA20}
Marco Fanizza, Farzad Kianvash, and Vittorio Giovannetti.
\newblock Quantum flags and new bounds on the quantum capacity of the
  depolarizing channel.
\newblock {\em Phys. Rev. Lett.}, 125:020503, Jul 2020.
\newblock \doi{10.1103/PhysRevLett.125.020503}.

\bibitem{Reichardt10}
Ben~W Reichardt.
\newblock Least span program witness size equals the general adversary lower
  bound on quantum query complexity.
\newblock Technical Report TR10-057, Electronic Colloquium on Computational
  Complexity, Apr 2010.
\newblock \urlprefix\url{https://eccc.weizmann.ac.il/report/2010/075/}.

\bibitem{Reichardt10a}
Ben~W Reichardt.
\newblock Span programs and quantum query algorithms.
\newblock Technical Report TR10-110, Electronic Colloquium on Computational
  Complexity, Jul 2010.
\newblock \urlprefix\url{https://eccc.weizmann.ac.il/report/2010/110/}.

\bibitem{FawziSaundersonEA19}
Hamza Fawzi, James Saunderson, and Pablo~A. Parrilo.
\newblock Semidefinite approximations of the matrix logarithm.
\newblock {\em Foundations of Computational Mathematics}, 19(2):259--296, Apr
  2019.
\newblock \doi{10.1007/s10208-018-9385-0}.

\bibitem{LeditzkyLeungEA18a}
Felix Leditzky, Debbie Leung, and Graeme Smith.
\newblock Quantum and private capacities of low-noise channels.
\newblock {\em Phys. Rev. Lett.}, 120:160503, Apr 2018.
\newblock \doi{10.1103/PhysRevLett.120.160503}.

\end{thebibliography}
\end{document}